# Interpretable machine learning applied to on-farm biosecurity and porcine reproductive and respiratory syndrome virus

## Running title: Application of machine learning to pig farm biosecurity

Abagael L. Sykes[1], Gustavo S. Silva[2], Derald J. Holtkamp[2], Broc W. Mauch[2], Onyekachukwu Osemeke[2], Daniel C.L. Linhares[2], Gustavo Machado[1]*

[1] Department of Population Health and Pathobiology, College of Veterinary Medicine, Raleigh, North Carolina, USA.
[2] Veterinary Diagnostic and Production Animal Medicine Department, College of Veterinary Medicine, Iowa State University, Ames, Iowa, USA.

**Corresponding author:** gmachad@ncsu.edu

## Summary

Effective biosecurity practices in swine production are key in preventing the introduction and dissemination of infectious pathogens. Ideally, on-farm biosecurity practices should be chosen by their impact on bio-containment and bio-exclusion, however quantitative supporting evidence is often unavailable. Therefore, the development of methodologies capable of quantifying and ranking biosecurity practices according to their efficacy in reducing disease risk have the potential to facilitate better informed choices of biosecurity practices.

Using survey data on biosecurity practices, farm demographics, and previous outbreaks from 139 herds, a set of machine learning algorithms were trained to classify farms by porcine reproductive and respiratory syndrome virus status, depending on their biosecurity practices and farm demographics, to produce a predicted outbreak risk. A novel interpretable machine learning toolkit, *MrIML-biosecurity*, was developed to benchmark farms and production systems by predicted risk, and quantify the impact of biosecurity practices on disease risk at individual farms.

Quantifying the variable impact on predicted risk 50% of 42 variables were associated with fomite spread while 31% were associated with local transmission. Results from machine learning interpretations identified similar results, finding substantial contribution to predicted outbreak risk from biosecurity practices relating to: the turnover and number of employees; the surrounding density of swine premises and pigs; the sharing of haul trailers; distance from the public road; and farm production type. In addition, the development of individualised biosecurity assessments provides the opportunity to better guide biosecurity implementation on a case-by-case basis. Finally, the flexibility of the *MrIML-biosecurity* toolkit gives it the potential to be applied to wider areas of biosecurity benchmarking, to address biosecurity weaknesses in other livestock systems and industry relevant diseases.

**Introduction**

Porcine reproductive and respiratory syndrome virus (PRRSV) is the most economically relevant endemic disease for the North American swine industry (Neumann et al., 2005; Holtkamp et al., 2013; Pileri and Mateu, 2016), and widely endemic in Europe (Zimmerman et al., 2019; Renken et al., 2021). PRRSV infection is associated with reproductive losses, and reduced growth performance from birth to market (Pileri and Mateu, 2016; Renken et al., 2021), with an estimated cumulative incidence in U.S. breeding herds of between 20-40% from 2010 to 2014 (Tousignant et al., 2015), and an estimated economic burden of more than $664 million/year in the U.S. (Holtkamp et al., 2013), and between €100 and €570 per sow in Europe (Renken et al., 2021).

Unfortunately, recent efforts directed at regionally controlling or eradicating PRRSV in the U.S. have not been fully successful, which may be attributed, in part, to low enrollment in such projects and a lack of understanding of regional pig dynamics (Corzo et al., 2010; Valdes-Donoso et al., 2016). In contrast, countries such as Sweden, Norway, and Switzerland have successfully controlled the virus through total depopulation/repopulation strategies, and other European countries such as Denmark controlling PRRSV with a combination of biosecurity measures and immunization strategies (Baekbo and Kristensen, 2015; Rathkjen and Dall, 2017). Whilst effective biosecurity practices may lead to significant improvements in productivity (Rodrigues da Costa et al., 2019; Kruse et al., 2020), in some cases they can be cost-prohibitive. Therefore, when choosing biosecurity practices, producers and veterinarians commonly balance their effectiveness against pathogen transmission with cost; however, the most effective practices are not necessarily the most economically efficient (i.e., depopulation, farm closure) (Corzo et al., 2010; Pileri and Mateu, 2016; Nathues et al., 2018; Jurado et al., 2019; Silva et al., 2019).

Fortunately, the selection and implementation of biosecurity practices can be facilitated by the use of biosecurity assessments, which are often able to highlight biosecurity practices with potential for improvement (Holtkamp et al., 2012; Gelaude et al., 2014; Silva et al., 2018, 2019; Rodrigues da Costa et al., 2019; Sasaki et al., 2020; Alarcón et al., 2021). While traditionally this are performed through rigorous but time-consuming surveys (Rodrigues da Costa et al., 2019; Alarcón et al., 2021), recent developments in swine health management software, such as Biocheck.UGent (Gelaude et al., 2014; Ghent University, 2021), BioAsseT (Sasaki et al., 2020), and ASF combat (Boehringer Ingelheim, 2018), have allowed the incorporation of biosecurity information to guide the implementation of biosecurity practices, through weight based analyses. Alas, these analyses are prone to subjectivity and often precluded by the lack of quantitative measures which could provide insight on the impact of on-farm biosecurity practices on infection risk (Silva et al., 2019; Jara et al., 2020; Alarcón et al., 2021; Galvis et al., 2021).

With the recent advancements of interpretable machine learning that allows for an improved understanding of model reasoning (Barredo Arrieta et al., 2020; Lucas, 2020; Molnar, 2021), there has been an emergence of research into Explainable AI (XAI); which has led to an increased development of machine learning methodologies with more explainability (Yang et al., 2021). By offering accessible and interpretable explanations regarding machine learning mechanisms, XAI allows researchers to better understand and better apply the insights gained from their models. In swine health management, there is a clear potential to apply such explainable and interpretable methodologies to on-farm biosecurity to help overcome the lack of quantitative analysis and benchmarking applied to biosecurity assessments (Silva et al., 2019; Kruse et al., 2020; Alarcón et al., 2021). Such data-driven approaches allow for straightforward benchmarking, including risk comparison between farms and quantification of the impact of biosecurity practices on the risk of new pathogen introduction and propagation (Fountain-Jones et al., 2019; Silva et al., 2019; Lucas, 2020; Biecek and Burzykowski, 2021; Ezanno et al., 2021; Fountain-Jones et al., 2021; Molnar, 2021). In particular, the application of local interpretation methods -- model agnostic techniques used to explain the predictions made by the model for individual data points (Carvalho et al., 2019; Molnar, 2021) -- may allow for the precise assessment and ranking of biosecurity practices by its importance to the infection risk of individual farms (Ribeiro et al., 2016; Lucas, 2020; Molnar, 2021). Furthermore, the ability to interpret relationships between biosecurity practices and disease outbreaks may be directly applied in the prevention of disease spread, through the prioritization of effective biosecurity measures and identification of common differences between farms experiencing outbreaks and those which are not (Silva et al., 2019; Neethirajan, 2020; Fountain-Jones et al., 2021).

In this study, we developed and applied an interpretable machine learning methodology to assess the impact of on-farm biosecurity practices on the predicted risk of PRRSV outbreaks. We achieved this through extending our previous interpretable machine learning framework and R package, *MrIML* (Fountain-Jones et al., 2020; Machado, 2021), creating *MrIML-biosecurity,* a new specialized machine learning toolkit. *MrIML-biosecurity* is capable of predicting and benchmarking PRRSV outbreak risk based on biosecurity practices and farm demographics through i) "global benchmarking", which allows for the identification of the most important variables and estimation of PRRSV predicted risk at both the farm and production system level[1]; and ii) "local benchmarking", which allows for the estimation of variable contribution to PRRSV risk predictions at a single farm.

**Material and Methods**

The data used in this study were collected from 139 breeding farms randomly provided by the 11 swine productions from 16 U.S. states who agreed to participate. These states were: Colorado, Illinois, Indiana, Iowa, Kansas, Minnesota, Missouri, North Carolina, North Dakota, Nebraska, Ohio, Oklahoma, South Dakota, Texas, Wisconsin and Wyoming. Each pig production system consented to share information regarding the number of PRRSV outbreaks from the previous

---

[1] Pig production systems are defined here as farms which are managed, owned or associated with the same company or integrator

five years, their biosecurity practices, farm location, and information about their neighborhood (i.e., number of farms within a three-miles radius; capacity of pigs within a three-mile radius) for a minimum of 10 premises. Data was collected through a single survey comprising 42 questions (Silva et al., 2019), implemented in 2019. PRRSV outbreaks were self-identified by farms through RNA-based diagnostic tests in pools of weaning age piglets and reported via the biosecurity survey. All PRRSV outbreaks, whether stable or unstable as described in (Holtkamp et al., 2011), were classed as a positive case. Further details of the data collection and the selection of the 42 variables have been previously described (Silva et al., 2019).

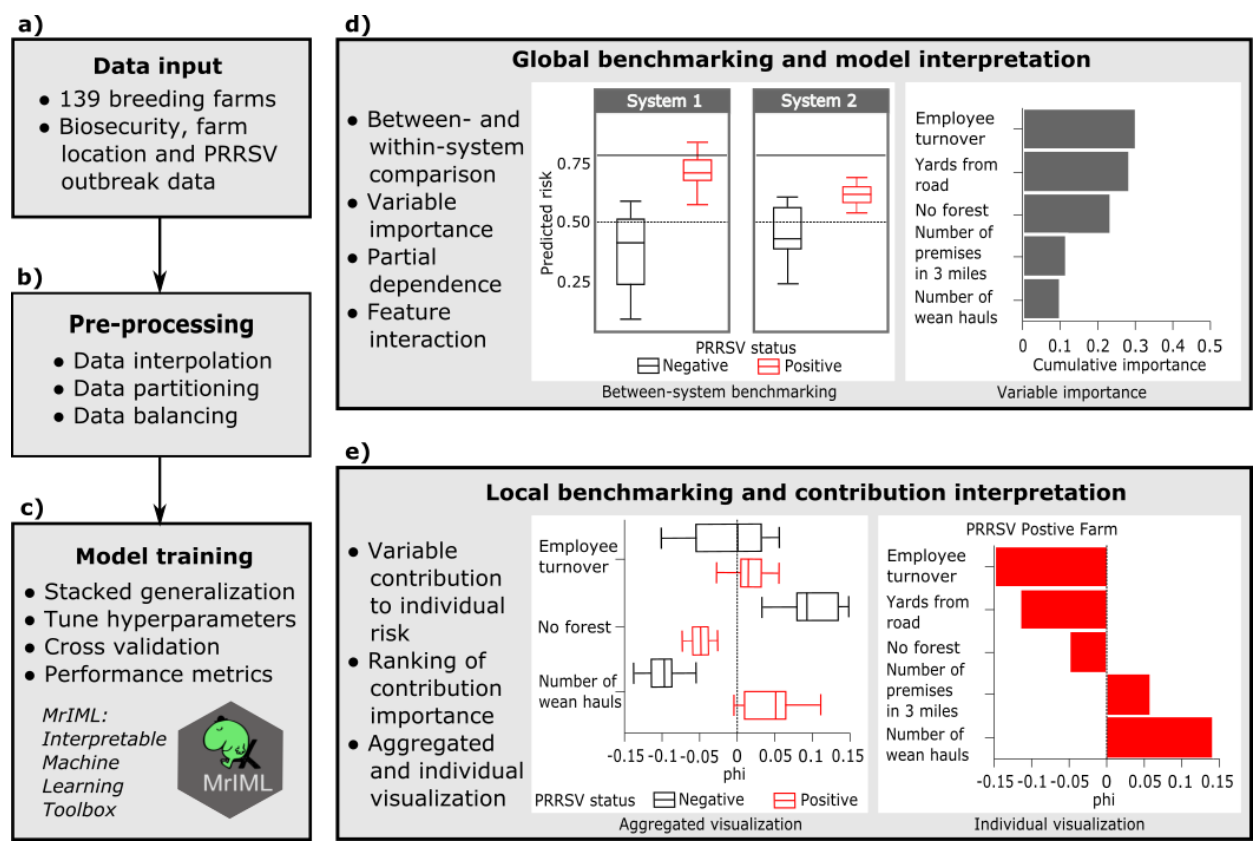

**Figure 1. Flow chart of the implemented machine learning pipeline and related outputs.** (a) Data regarding biosecurity practices and farm demographics and the number of PRRSV outbreaks in the past five years were collected via a 42 question survey (Silva et al., 2019). (b) Balancing techniques were implemented due to the uneven distribution of outcomes and followed by data partitioning to create test and training data. (c) Individual machine learning models and an ensemble model were trained, employing cross-validation techniques to reduce overfitting. Using performance metrics, the most appropriate model was selected and used to produce all results. (d) Global benchmarking consisted of predicted risk benchmarking among (left panel) and within systems, and model interpretation via variable importance (right panel), partial dependence profiles, and feature interaction. (e) Local benchmarking consisted of the interpretation and benchmarking of biosecurity practice and farm demographic contribution to the predicted PRRSV outbreak risk at individual farms, shown by the *phi* value, presented as aggregated summary box plots (left panel) and individualized waterfall plots (right panel).

*The machine learning pipeline*

The interpretable machine learning, *MrIML* R package, available in GitHub (Fountain-Jones et al., 2021; Machado, 2021), was expanded to allow for the stacked generalization of multiple algorithms (Breiman, 1996). In addition, new functionalities were developed under the *MrIML-biosecurity* R package module, to allow for: i) "global benchmarking", which encompasses the identification of the most important biosecurity practices and farm demographics, and estimation of PRRSV predicted risk at both the production system and farm level; and ii) "local benchmarking", which includes the estimation of biosecurity practice and farm demographic contribution to PRRSV risk predictions at a single farm. Innovatively, as part of the local benchmarking functionality, biosecurity practice and farm demographic contributions were aggregated and ranked to summarize their impact at a local level (Figure 1).

*Model implementation*

Within the proposed machine learning framework we considered three candidate algorithms: Support Vector Machine (SVM) (Boser et al., 1992); Random Forest (RF) (Breiman, 2001); and Gradient Boosting Machine (GBM)(Friedman, 2001). Each algorithm was tuned to balance over and underfitting, and trained using 75% of the dataset, labeled as training data, and evaluated against the remaining 25% of the dataset, labeled as test data (Figure 1). These datasets were resampled 10 times from the original dataset through 10-fold cross-validation, randomly splitting the data into equal size folds (Machado et al., 2019). Further details of model tuning are presented in the Supplementary Material Section "Model tuning" and the *MrIML-biosecurity* vignette (Machado, 2021). In addition, undersampling methodologies were employed to balance the data due to unequal distribution of the outcome, the presence or absence of historical cases of PRRSV. Undersampling methods have been proposed as a good means of increasing the sensitivity of a classifier to the less represented outcome, referred to as the minority class (Machado et al., 2019; Silva et al., 2019). In *MrIML* we provide wrappers for one down-sampling and two common up-sampling routines: (1) Synthetic Minority Over-sampling Technique (Chawla et al., 2002) and (2) Random Over-Sampling Examples (Lunardon et al., 2014). In this study, we evaluated the above three approaches, where downsampling presented the best performance and was subsequently used to balance the study dataset.

A model consensus was created from the above candidate machine learning algorithms through a stacked generalization approach (Polikar, 2012). Stacking is a regularized linear model approach that can reduce the generalization error by training multiple primary learning algorithms and combining their predictions (Dietterich, 2000; Polikar, 2012). Here we used the random forest as the base model to which the SVM and GBM algorithms were stacked. The performance of the stacked model was then compared against the individual algorithms, using

Matthews correlation coefficient (MCC), the area under the curve (AUC), sensitivity, and specificity. For more details on these metrics see Supplementary Material Section "Performance metrics". The best performing algorithm or ensemble was re-applied to the full dataset to produce predicted PRRSV outbreak risk values.

*Global benchmarking: Model interpretation*

Variable importance, partial dependence, and feature interaction were generated for the best performing model to interpret the overall impact of biosecurity practices and farm demographics on the risk of PRRSV outbreaks. Here, variable importance can be defined as the dependence between PRRSV outbreak risk prediction, and the biosecurity practices and farm demographics. In this study variable importance was implemented through the Gini Index, which ranks variables according to their upscaled node impurity (Wright et al., 2016) (Figure 2). The top four biosecurity practices and demographics ranked here were further analyzed through partial dependence profiles (Friedman, 2001) (Figure S1 to S11). Partial dependence profiles are a methodology that allows the direct visualization of the relationship between values of a predictor variable, such as a biosecurity practice or farm demographic, and the outcome, such as predicted risk values, after accounting for other predictors (Friedman, 2001; Elith et al., 2008; Molnar, 2021). Finally, to detect interactions between biosecurity practices and farm demographics, interaction strengths across all pairs of variables were calculated through Freidman's H statistic (Friedman and Popescu, 2008) and ranked according to their importance, via the "*mrInteractions*" function of *MrIML* (Biecek and Burzykowski, 2021; Fountain-Jones et al., 2021; Molnar, 2021) (Figure S12).

*Global benchmarking: Comparison of predicted risk of PRRS outbreak among and within production systems*

To further interpret risk, a discretization methodology (Kuhn and Wickham, 2021) was applied to predicted PRRSV risk values, which used percentile breaks to generate three equally sized risk categories: low, medium, and high-risk. Using the *MrIML-biosecurity* function "*mrBenchmark*", production systems were benchmarked via their distribution of predicted outbreak risk for farms reporting outbreaks and farms with no reported outbreaks (Figure 3). In addition, "*mrBenchmark*" was used to also benchmark each farm within the production system (Figure S13). Finally, the density distributions of biosecurity practices and farm demographics within each risk category were presented (Figure 4 and Figures S14 to S23).

*Local benchmarking: Quantifying biosecurity practice and farm demographic contribution at individual farms*

To identify which biosecurity practices and farm demographics contributed the most to predicted PRRSV outbreak risk at individual farms, a local model-agostic explainer, *breakDown* (Staniak and Biecek, 2019), was implemented via the *MrIML-biosecurity* function "*mrLocalExplainer*". Briefly, *breakDown* is used to explain the influence of individual variables, such as biosecurity practices and farm demographics, over the outcome or prediction at a single data point, such as an individual farm. "*mrLocalExplainer*" implemented the *breakDown* model via the "broken" function of the *breakDown* package (Staniak and Biecek, 2019), which identifies variables in the data instance, *x*, which cannot be altered without a change in the prediction, *f(x)*. *breakDown* achieves this by searching for the combination of variables, *vars*, which will change the value of the prediction to its expected value, *E[f(x)]*. The distance between these predictions, presented in equation 1 below, can be used to calculate the contribution, *phi*, of each variable on the change in prediction:

$$d(f^{\ vars}(x), f(x)) \ = \ f^{\ vars}(x) \ - \ f(x) \qquad (1)$$

The size of *phi* can be further interpreted as the importance of the variable to the predicted outbreak risk at data instance *x* (Staniak and Biecek, 2019). Briefly, biosecurity practices and farm demographics with a *phi* > 0 contribute to an increase in predicted PRRSV outbreak risk , while practices and demographics with a *phi* < 0 contribute to a decrease in predicted outbreak risk. The overall distribution of these individual contribution values across farms reporting outbreaks and farms with no reported outbreaks was summarized as box plots (Figure 5 and Figure S24). We then further explored these contributions, by plotting *phi* values against the observed values of the practice or demographic in question, to understand how contribution varies as variable values change (Figures S25 to S35). Finally, we presented individual biosecurity practice and farm demographic contributions as waterfall plots for two example farms (Supplementary Figure S36 and S37).

**Results**

Overall, 116 (83.5%) farms were commercial sites and 23 (16.5%) were genetic multiplication sites (primarily raising genetic breeding replacements). In total, 89 farms reported at least one outbreak in the last five years, and the remaining 50 reported no outbreaks during this time frame. Of the farms reporting outbreaks, 89.9% reported between one and three outbreaks (Figure S38). Descriptive results regarding the distribution of farms and PRRSV outbreaks, and a summary of the biosecurity practices and farm demographics can be found in Supplementary Figures S38, S39 and S40, and Supplementary Tables S1-S3, respectively.

*Model performance*

Due to the imbalance between farms reporting at least one outbreak and farms with no reported outbreaks, model performance was compared between the original dataset and a dataset balanced via downsampling. In three out of four performance metrics, the model trained with the unbalanced dataset out-performed the model trained with the downsampled dataset. Higher values of AUC (0.93), and specificity (92%) indicate better diagnostic accuracy of the unbalanced model, while a higher MCC (0.72) indicates a better correlation between the observed outcomes in the data and predicted classifications.

*Global benchmarking: Model interpretation*

The classification of farms and prediction of PRRSV outbreak risks were primarily driven by biosecurity practices rather than farm demographics, though location of other farms and distance to the public road appeared to be particularly important demographics (Figure 2). The top four biosecurity practices and farm demographics identified during variable importance were: i) raising genetic replacements; ii) yards to the nearest public road; iii) capacity of breeding females within a three-mile radius, and iv) annual turnover of on-site employees (averaged from the previous 2 years). In contrast, the least important practices and demographics were: i) topography of the surrounding area; ii) capacity of boars within a three-mile radius; iii) the number of dead animal removals per month, and iv) capacity of show pigs within a three-mile radius.

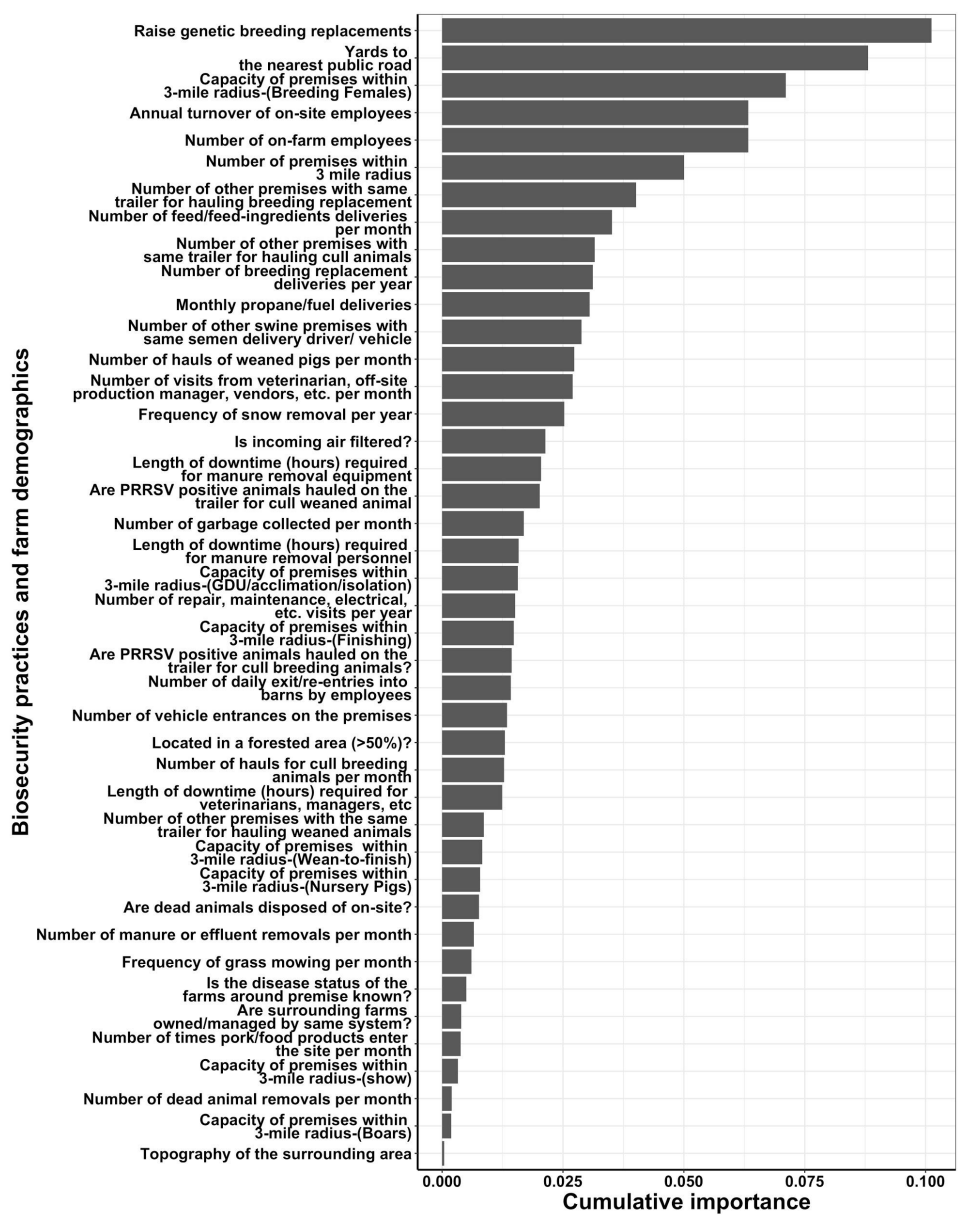

**Figure 2. Cumulative importance of biosecurity practices and farm demographics.** The y-axis is (top to bottom) in descending order of importance. Cumulative importance values calculated through the Gini Index are represented by the x-axis. Higher cumulative importance values indicate a higher order of importance in relation to PRRSV outbreaks.

The relationships between predicted PRRSV outbreak risks and the individual variables were further explored through partial dependence profiles in Figures S1 to S11. Farms raising genetic breeding replacements (multiplication sites) appear to have a clear reduction in predicted PRRSV outbreak risk compared to farms raising commercial pigs (commercial sites) (Figure S1A). There was also evidence of a clear linear relationship between distance to the nearest public road and PRRSV outbreak risk, with predicted risk decreasing as distance from the road increases, eventually plateauing above approximately 700 yards (Figure S1B). In contrast, as the capacity of breeding females in the surrounding three miles increases, there is a notable increase in predicted PRRSV outbreak risk (Figure S1C). However, above a capacity of 8000 breeding females, the predicted risk appears to reduce again indicating a potential non-linear relationship between the surrounding capacity of breeding females and predicted risk. Similarly, there appears to be a non-linear relationship between annual turnover of on-site employees and predicted PRRSV outbreak risk, with an initial decrease in predicted risk until an employee turnover of 35%, above which the predicted risk begins to increase (Figure S1D).

Additionally, interactions within the biosecurity practices and farm demographics were investigated. The top interactions were between raising genetic replacements and: PRRSV positive animals being hauled on the same trailer used for weaned animals; monthly propane/fuel deliveries; the number of daily entries and exits into the barns by employees; the number of on-site employees; and the number of other premises using the same trailer for hauling breeding replacements (Figure S12).

*Global benchmarking: Comparison of predicted risk among and within production systems*

Using a discretization method to generate risk categories, farms with PRRSV predicted risk values below 0.46 were considered low risk, between 0.46 and 0.86 were considered medium risk and above 0.86 were considered high risk. These thresholds were then used to benchmark the predicted risk among production systems (Figure 3). Briefly, farms of systems one, six, 10 and 11 appear to be at a heightened risk, as the majority of farms reporting outbreaks in these systems have predicted risk values that are well above the high-risk threshold, suggesting that those sites are very likely to experience continued PRRSV outbreaks (Figure 3). In contrast, system seven seems to have a lower overall risk compared to other production systems, as the majority of farms reporting no outbreaks are below the low-risk threshold. In addition, the farms reporting at least one outbreak have a much lower predicted risk compared to similar farms of other systems, suggesting that new PRRSV outbreaks are less likely to occur in this system (Figure 3). Additionally, an example of within-system benchmarking is presented in Figure S13. In the example, the majority of farms reporting no outbreaks are within the low-risk category while the majority of farms reporting at least one outbreak are within the medium and high-risk categories (Figure S13), suggesting that farms with no reported outbreaks in this system are less likely to experience outbreaks compared to farms with at least one reported outbreak.

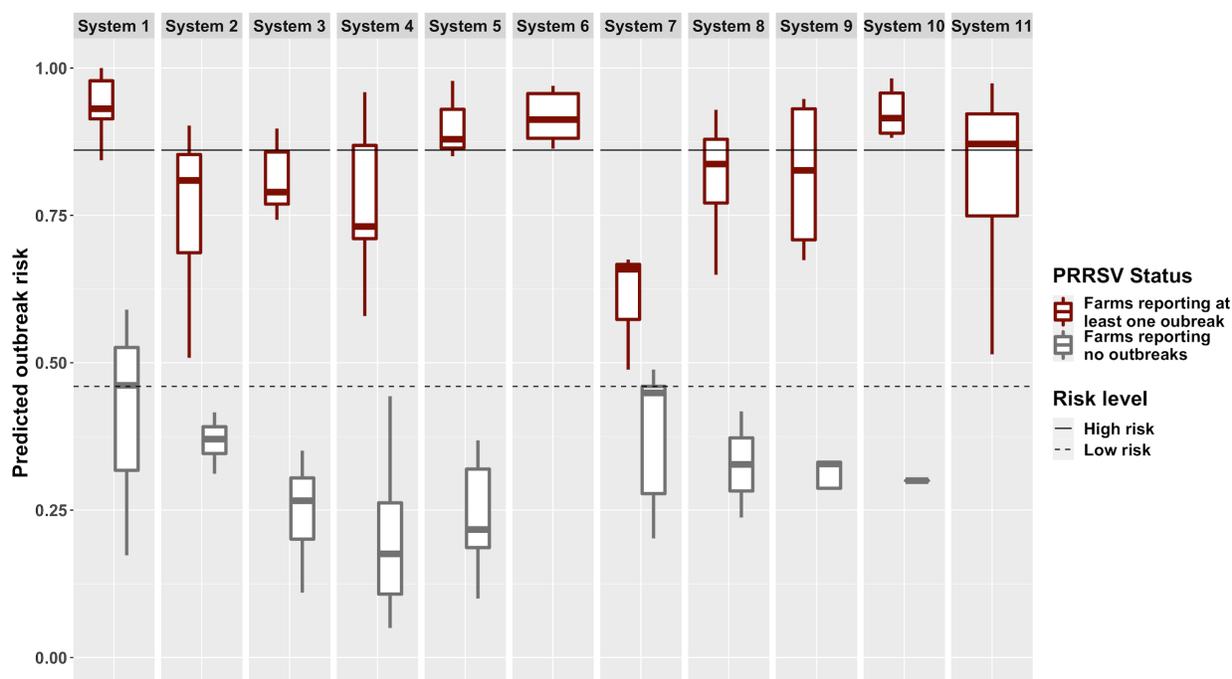

**Figure 3. Distribution of farms reporting PRRSV outbreaks (none versus at least one) by production system in relation to the risk of a new outbreak.** The x-axis represents the production system identification and the y-axis represents the predicted PRRSV outbreak risk produced by the machine learning model. Farms below the dashed line are considered low risk, between the dashed line and the solid line are considered medium risk, and above the solid line are considered high risk.

Finally, we described the distribution of the biosecurity practices and farm demographics in relation to the three defined risk categories (Figure 4 and Figures S14 to S23). The majority of multiplication sites were in the low-risk category, while all farms in the high-risk category and the majority of farms within the medium-risk category were commercial sites (Figure 4A). As for the distance from the public road, the majority of farms in all risk categories were between 0 and 500 yards from the public road, however, farms in the high-risk category appear to peak in number at shorter distances to the road, compared to the low and medium-risk categories (Figure 4B). In contrast, the capacity of breeding females in the surrounding three miles appears to be much lower in the low and medium-risk categories, with most farms having between 0 and 5000 breeding females in the area, compared to the high-risk category in which most farms have between 0 to 10000 breeding females in the area (Figure 4C). Finally, for annual turnover, all three risk categories peaked around 35%, however the high-risk category also had a significant number of farms with a turnover proportion of up to 100% (Figure 4D). See Supplementary Figures S8 to S17 for the description of all other biosecurity practices and farm demographics.

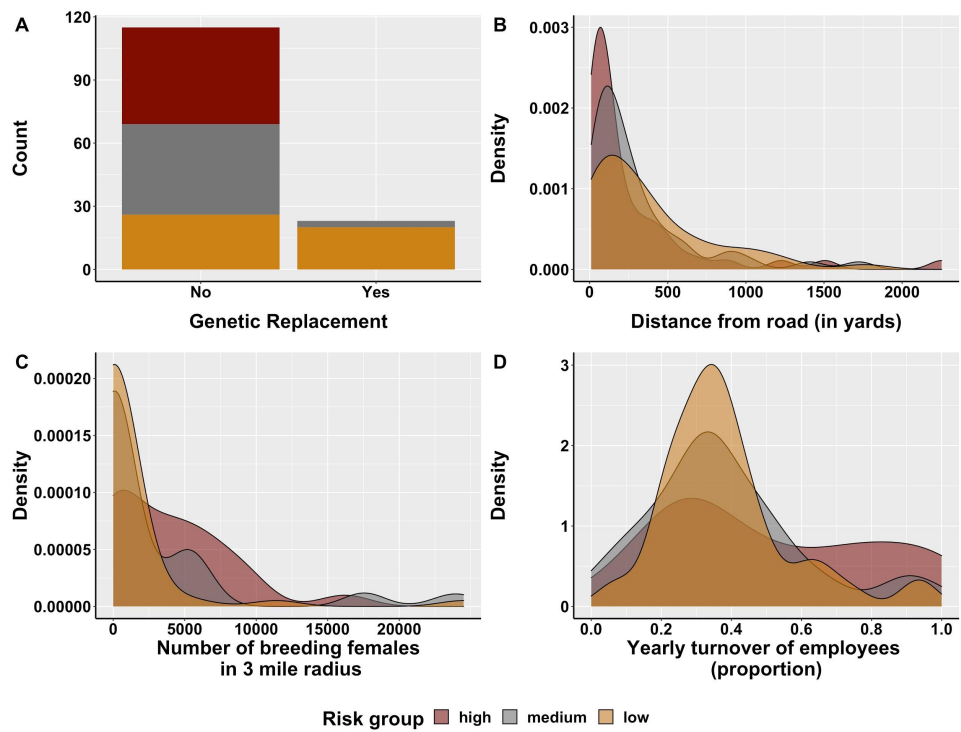

**Figure 4. Distribution of observed biosecurity practice and farm demographic values by PRRSV outbreak risk category.** The predicted PRRSV outbreak risk values were categorized by a *tidymodels* discretization method to create categories for low, medium, and high predicted outbreak risk (Kuhn and Wickham, 2021). (A) distribution of farms raising genetic breeding replacements (p < 0.000); (B) distribution of farm distance from the public road (p = 0.37); (C) distribution of the number of breeding females in the surrounding three-miles (p < 0.05); and (D) distribution of annual turnover of employees shown via a proportion (0% to 100%) (p < 0.05).

*Local benchmarking: Quantifying biosecurity practice and farm demographic contribution at individual farms*

The *breakDown* explainer was used to identify and rank the contribution of biosecurity practices and farm demographics to the predicted PRRSV outbreak risk at individual farms. Using "*mrLocalExplainer*", the *phi* values for farms reporting at least one outbreak and farms with no reported outbreaks were aggregated and used to rank variables by their prediction reliability across all farms (Figure 5 and Figure S24). Here, the top four ranked biosecurity practices which had the largest contribution to PRRSV risk at individual farm-level were: i) the number of other premises using the same trailer to haul breeding replacements ii) the number of daily on-site employees; iii) the number of other premises with the same trailer for hauling cull animals; and iv) the annual turnover of on-site employees (Figure 5). Within the ranked farm demographics, the variables with the greatest contribution to predicted PRRSV outbreak risk at individual farm-level were: i) capacity of breeding females within the surrounding three-miles; ii) whether the premise is primarily raising genetic replacements; iii) yards to the nearest road; and iv) and the number of premises in the surrounding three-miles (Figure S24).

Following the aggregate ranking of variables, "*mrLocalExplainer*" was used to explore the directionality of contribution for all biosecurity practices and farm demographics, to investigate the change in contribution as the observed values of the variables change. For simplicity, we will only discuss the results for the top biosecurity practices identified within the aggregated *breakDown* plot (Figure S25). Individually, lower numbers of farms using the same trailer for hauling breeding replacements (Figure S25A) and lower numbers of on-site employees (Figure S25B) appear to contribute to a decrease in predicted risk. However, as these values rise above five and 10, respectively, they begin to contribute towards an increase in predicted risk (Figure S25A and Figure S25B, respectively). In contrast, as the number of other premises using the same trailer for hauling cull animals increases above two, there is a substantial contribution to a decreased predicted risk (Figure S25C). Finally, the annual turnover of on-site employees appears to have minimal contribution to predicted risk at lower proportions, however as turnover increases past 50%, there is an increasing contribution to increased predicted risk (Figure S25D). Supplementary Material Figures S26 to S35, present the directionality of local contribution for the remaining biosecurity practices and farm demographics.

Under the local benchmarking functionality, our *MrIML* package, vignette (https://nfj1380.github.io/mrIML/articles/Vignette_biosecurity.html) and shiny app (https://machado-lab.github.io/Software/) we also produced reliable individualized contribution ranking at each farm, examples of which are presented in Figure S37 and S38. Briefly, Figure S37 presents the results of a farm with at least one reported outbreak, in which having zero farms and zero breeding females in the surrounding three miles contributes towards a lower predicted risk. In contrast, having an annual turnover proportion of 90% and being situated only 42 yards from the main road appears to contribute to an increased predicted outbreak risk. Similarities can be seen within the farm with no reported outbreaks presented in Figure S38. Consistently, having zero farms and zero breeding females in the surrounding three miles is contributing to a decreased predicted risk, while sharing a semen delivery vehicle with six other premises and not raising genetic replacements (i.e., a commercial site) is contributing to an increased predicted outbreak risk.

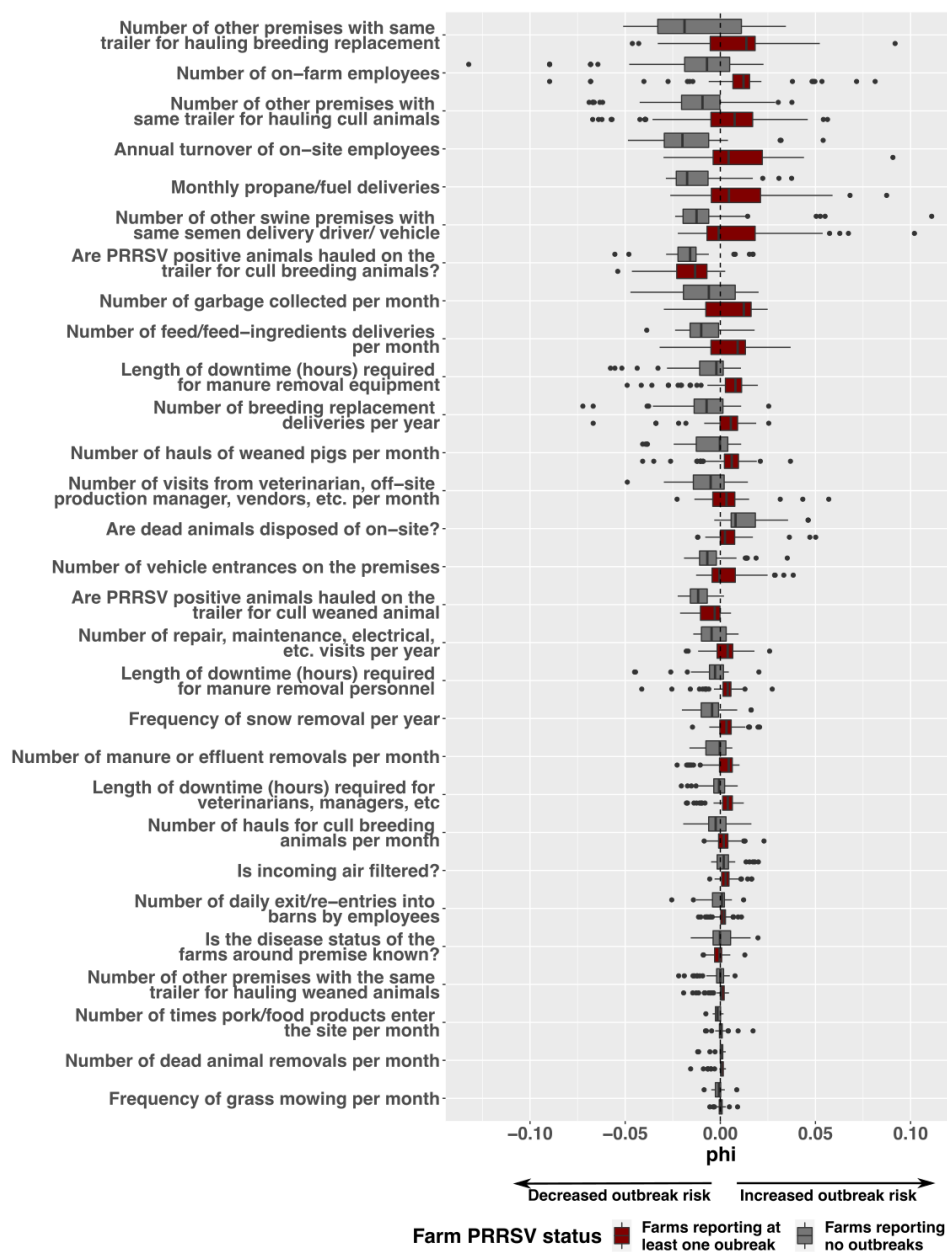

**Figure 5. The rank of biosecurity practices produced by local model agnostic interpretation methods.** The boxplots represent a summary of the breakDown values generated for each biosecurity practice per farm, ordered by mean *phi* value. The current status of farms is presented having reported at least one outbreak (red) or having reported no outbreaks (gray). The axis represents the contribution of the practice to PRRSV outbreak prediction. Values above zero are contributing to an increased risk prediction and values below zero are contributing to a decreased risk prediction.

**Discussion**

Through *MrIML-biosecurity*, a specialized branch of the interpretable machine learning toolbox in the *MrIML* R package (Fountain-Jones et al., 2021), we developed and implemented a novel interpretable machine learning methodology. The *MrIML-biosecurity* methodology compares predicted PRRSV outbreak risk among and within pig production systems, and ranks on-farm biosecurity practices and farm demographics by their impact on predicted risk. We demonstrated significant agreement in the directionality of PRRSV outbreaks risk between average partial dependence plots (Figures S1 to S11) and aggregated *breakDown* results (Figures S25 to S35), identifying the variables most relevant to reporting a new PRRSV outbreak as: raising genetic breeding replacements; the surrounding density of farms and breeding females; the sharing of haul trailers for breeding replacements and cull animals; the annual turnover and number of on-site employees; and distance to the public road. Quantifying the relative effect of on-farm biosecurity practices on the probability of new outbreaks demonstrated that up to 50% of features were associated with fomites, such as sharing haul trailers and other vehicles, while 31% of features were associated with local transmission (i.e. aerosol and mechanical transmission), such as a high surrounding density of swine farms and pig capacity (Otake et al., 2002, 2003, 2010; Dee et al., 2004, 2009; Otake et al., 2004; Pitkin et al., 2009; Arruda et al., 2019; Jara et al., 2020; Galvis et al., 2021). Our new methodology demonstrated the potential application of the interpretable machine learning framework to support veterinarians and production companies in identifying key biosecurity aspects associated with PRRSV outbreaks in breeding herds. More importantly, this is the first proposed methodology that allows for individual farms to identify their most relevant biosecurity practices, thus allowing individualized biosecurity plans to be developed in a case-by-case fashion ([https://nfj1380.github.io/mrIML/](https://nfj1380.github.io/mrIML/)).

Previous investigations have identified associations between the access and sharing of vehicles and employee movements, with an increased risk of PRRSV infection (Evans et al., 2008; Lambert et al., 2012; Silva et al., 2019; Black et al., 2021). This is in support of our findings in which increases in sharing of trailers for hauling breeding replacements and increases in the number of on-site employees contributed to an increased PRRSV outbreak risk (Figure S25). On the other hand, we identified associations between decreases in predicted PRRSV risk and increases in the number of farms sharing the same trailer for hauling cull animals (Figure S25). This was surprising as we would expect to observe an increase in risk as access to the premise by vehicles increases, due to the rising potential for the introduction of PRRSV via fomites (Romagosa, 2017; Black et al., 2021). However, it is possible that the presence of a high-risk practice such as sharing haul trailers increases the producer awareness of disease risk and influences the implementation of other biosecurity practices to mitigate possible disease introduction (Merrill et al., 2019; Pudenz et al., 2019; Lee et al., 2021).

In addition to the number of on-site employees, employee turnover was also identified as a particularly important variable in the prediction of the PRRSV outbreak risk, with predicted risk increasing as the proportion of annual turnover increases. High turnover rates are a widespread and well known issue in the U.S. swine industry due to changing rural demographics and immigration policies which affect the labor supply (Boessen et al., 2018). Biosecurity protocols utilized in the swine industry are often complex and must be completed in sequence, which could lead to a lack of biosecurity compliance in new untrained or inexperienced employees (Racicot et al., 2012). This has been discussed previously by Racicot et al., (2012), who identified an association between a lack of experience in workers and a decrease in biosecurity compliance. It was suggested that individuals with a lack of experience may not be fully aware of the consequences of non-compliance and therefore may not perceive the importance of following biosecurity protocols (Racicot et al., 2012; Rabinowitz et al., 2013). These results highlight the substantial impact of high turnover in the swine industry on the health, and consequently the productivity, of swine production systems. Concerningly, we found an association of turnover with particular production systems, which may indicate systems which may be particularly vulnerable to biosecurity non-compliance (Racicot et al., 2012; Rabinowitz et al., 2013). However, we hasten to acknowledge that resolving high turnover is a difficult challenge requiring inter-sectoral cooperation between government and industry, and further research into the underlying factors leading to voluntary and involuntary turnover is required (Racicot et al., 2012; Rabinowitz et al., 2013; Boessen et al., 2018).

Furthermore, we identified a substantial contribution of farm demographics to an increased predicted PRRSV outbreak risk, primarily driven by decreases in the proximity of farms to the public road, and increases in the density of surrounding swine farms and the surrounding capacity of breeding females. Previous studies have identified similar demographic variables as risk factors for the local transmission of PRRSV, which includes mechanical and aerosol transmission (Mortensen et al., 2002; Evans et al., 2008; Lambert et al., 2012; Velasova et al., 2012; Arruda et al., 2019; Jara et al., 2020; Galvis et al., 2021). Several experimental studies have succeeded in producing results to support the hypothesis of local transmission, showing viable PRRSV in air samples up to 9.7 km from an infected herd and in the guts of houseflies (Otake et al., 2003; Otake et al., 2004; Otake et al., 2010; Dee et al., 2009; Arruda et al., 2019). However, there are still questions regarding the effectiveness of local transmission under field conditions (Arruda et al., 2019). Moreover, the farm demographics identified here are unfeasible to alter, highlighting a need to identify biosecurity practices which could be adopted to mitigate local spread instead (Galvis et al., 2021).

Another contributing farm demographic identified as significantly important, was whether the farm primarily raised genetic breeding replacements, also referred to as multiplication sites, or whether they raised commercial pigs, referred to as commercial sites. There is often a heavier emphasis on effective biosecurity at genetic multiplication sites due to the large number of farms they supply with replacement breeding animals, which would be at risk if an outbreak occurred (FAO et al., 2010; Ramirez and Zaabel, 2012; Pudenz et al., 2019).

While we observed a notable contribution of this demographic variable overall to an increased predicted risk (Figure 5), there was little difference in PRRSV outbreak risk when multiplication sites were compared with commercial sites (Figure S27). However, it is worth noticing that we found interactions between raising genetic breeding replacements and many other biosecurity practices and farm demographics, which could be masking unmeasured risk factors (Figure S12). When we consider only the marginal effect of the variable in the partial dependence plots (Figure S2), we found a much clearer difference in the dependence between multiplication and commercial sites. To further explore the association of production type and predicted PRRSV outbreak risk, there is a need for the continued development of improved machine learning methodologies that are capable of accounting for such interactions explicitly within the modelling pipelines.

We observed good agreement between global and local interpretation methods, with similarities in the top biosecurity practices and farm demographics identified between variable importance and the aggregated *breakDown* plots (Figure S24). Both interpretation methods can be advantageous, however, local methods provide the opportunity to individualize biosecurity assessments, capturing nuances in variable contribution which may not be detected in the global importance (Goldstein et al., 2015; Greenwell, 2017; Kopitar et al., 2019; Molnar, 2021). The use of local explanation methods in this study is in line with recent work in the field of explainable machine learning, to improve the explanation of model reasoning to make it understandable by stakeholders and other participants without the need for machine learning training (Ghai et al., 2021). Although local explainable machine learning is a relatively new approach to veterinary epidemiology, local interpretation methods have previously been applied to the agricultural sector to quantify the importance on of hydro-climatic factors on crop evapotranspiration, detect estrus in cattle, and predict peach fruit ripeness (Fauvel et al., 2019; Ljubobratovic et al., 2020; Chakraborty et al., 2021).

In addition to supporting an individualized approach to biosecurity practices, the machine learning methodology presented here also allows for the benchmarking of predicted outbreak risk among and within pig production systems. This allows production systems to assess performance compared to their peers and identify farms within their systems that are particularly at risk of PRRSV outbreaks. Benchmarking of farms according to their biosecurity practices is not a novel application; in 2011 the American Association of Swine Veterinarians released their Production Animal Disease Risk Assessment Program (PADRAP) to assess a farm's risk of the clinical outbreak by their biosecurity strategy (Holtkamp et al., 2012). Similar tools have been developed by several institutions and organizations including Biocheck.UGent at Ghent University (Gelaude et al., 2014); the ASF focussed tool "ASF Combat" at Boehringer Ingelheim (Boehringer Ingelheim, 2018); and BioAsset created by the PRRS-Japan Elimination team (Sasaki et al., 2020). However, these tools are built upon qualitative methodologies such as multi-criteria decision analysis and the assignation of weights or values to biosecurity practices based on expert opinion (Gelaude et al., 2014; Silva et al., 2018; Sasaki et al., 2020; Alarcón et al., 2021), which could potentially introduce bias into the risk estimates (Alarcón et al., 2021). In contrast, the machine learning-based methodology proposed in this study has no reliance on opinion. Instead, it uses only the data provided to quantify and predict risk and interpret complex and non-linear relationships which may not be captured otherwise (Silva et al., 2019). Additionally, as further farm data is collected, our approach will continue to develop its understanding of these relationships, improving its predictive performance (Rabinovich et al., 2021).

**Limitations and future directions**

Despite the advantages over parametric models in relation to predictive performance, there are several relevant limitations to the machine learning methodology utilized here (Elith et al., 2008; Machado et al., 2015; Lucas, 2020; Rabinovich et al., 2021) including the presence of interactions among variables which may not be identified and the lack of an independent study sample, potentially impacting model performance and leading to spurious interpretations (Strobl et al., 2009; Boulesteix et al., 2015; Wright et al., 2016; Oh, 2019). The implementation of methods such as interaction forests and pairwise importance techniques improve the identification of interactions in machine learning algorithms (Wright et al., 2016; Hornung and Boulesteix, 2021), while Individual Conditional Expectation (ICE) plots and the iBreakDown model are capable of accounting for known interactions in their post-hoc interpretations (Goldstein et al., 2015; Biecek and Burzykowski, 2021). This is an avenue of improved interpretation which should be explored in future studies to account for interactions.

Similarly, the use of partial dependence profiles (PDP) for global interpretation may not be ideal, given that PDP are based on averaging the variable effect which may lead to heterogeneous or small effects being overlooked and create a distribution for the variable of interest that is not accurate (Goldstein et al., 2015; Greenwell, 2017; Molnar, 2021). For example, in Figure S1C, which presents the partial dependence profile for the capacity of breeding females in the surrounding three miles, we observe a decrease in the association between the capacity of breeding females and a positive model prediction, after a capacity of approximately 8,000. However, there are only a small number of farms with more than 8,000 breeding females in the surrounding three miles, which makes the partial dependence average can be easily influenced by any outlying or anomalous results. Through the implementation of Individual Conditional Expectation (ICE) plots (Goldstein et al., 2015; Greenwell, 2017) it is possible to mitigate potential oversight; however, to fully observe small or individual effects, local interpretation methods are recommended (Molnar, 2021). Alas, our results should be interpreted with caution as local interpretation techniques have limitations regarding a large diversity of methodologies and lack of consensus about the use of specific approaches (Carvalho et al., 2019; Li et al., 2021). Fortunately, in a recent work, *MrIML* has used simulated data in which the ground truth of non-linear relationships between outcome and variables was known, and was able to fully explain the relationship between the predictors and the outcome of interest (Fountain-Jones et al., 2021).

Caution should also be taken in interpreting the results presented here, owing to the unrepresentative sample of herds included in the study. Unfortunately, due to time constraints only large industrial breeding farms were included in the data collection, reducing the external validity of the results, in particular their applicability to other industrial stages or smaller backyard herds. In the near future, we aim to expand the number and diversity of herds incorporated into the machine learning methodology. Additionally, the data collected may not accurately reflect the full biosecurity at the participating premises. Due to the number of farms from which we received data it was not feasible to visit every farm to ensure that biosecurity practices were being carried out in a consistent manner or whether premises were compliant with their biosecurity protocols, which could impact a premise's vulnerability to an outbreak (Mortensen et al., 2002; Racicot et al., 2012; Velasova et al., 2012; Rabinowitz et al., 2013;Vilata et al., 2020). Similarly we only collected biosecurity practice data at a single point in the study period and therefore were unable to verify that the biosecurity practices stated were implemented during each PRRSV outbreak, thus we can not rule out a non-compliance effect on the calculated PRRSV outbreak risk. Further studies could implement our developed methodology and collect further time series data to determine the impact of biosecurity measures on future outbreaks.

Other potentially confounding factors that were not considered in this model may include the use of PRRSV vaccinations, persistent PRRSV infections and comorbidities. While the modified live PRRSV vaccines available in the U.S. are surrounded by concerns regarding viral reversion and recombination with wild type strains, they do provide a notable level of protection which could reduce the risk of outbreak at premises (Liu et al., 2018; Eclercy et al., 2019; Montaner-Tarbes et al., 2019). In contrast the presence of persistent or recurrent PRRSV infections facilitating long term circulation within premises and comorbidities with other endemic diseases could increase the probability of a farm to experience an outbreak (Bierk et al., 2001; Rathkjen and Dall, 2017). While such data could be useful to incorporate when tailoring on-farm biosecurity, it increases the complexity of the model which may decrease the interpretability of the model's decisions (Carvalho et al., 2019).

Regardless of the limitations, the potential benefits of the methodology described here are wide-reaching. The data driven approach combined with a user-friendly interface (https://nfj1380.github.io/mrIML/articles/Vignette_biosecurity.html), has the potential to expand the use of quantitative methodologies to infer disease risk from on-farm biosecurity practices. Furthermore, we provide an online tool and a user-guide, based on this methodology, to assist veterinarians and farm managers in both commercial and independent mixed animal veterinarians to prioritize biosecurity practices and visualize their improvement in outbreak risk over time. Briefly, in the proposed interpretable machine learning tool (vignette and publicly available shiny app), veterinarians and managers can upload or update farm data for analysis and download the outputs to provide guidance when designing on-farm biosecurity plans. Additionally, the expansion of *MrIML* (Fountain-Jones et al., 2021) to include the *MrIML-biosecurity* R functions for global and local benchmarking, creates a practical and streamlined approach for the wider scientific community to apply a similar machine learning methodology to other areas of biosecurity research. Moreover, with the development of the MrIML package there is potential in the future to incorporate both genetic sequence data and multiple pathogens within a machine learning framework, allowing for the analysis of on-farm biosecurity in relation to multiple PRRSV strains and co-morbidities with other industry-relevant pathogens such as porcine epidemic diarrhea virus (PEDV), porcine circovirus 2 and 3 (PCV 2 and PCV 3), influenza A virus, classical swine fever virus (CSFV) and African swine fever virus (ASFV). This is especially pertinent for ASF, which is an imminent threat to the U.S. pork industry with potentially devastating consequences (Jurado et al., 2019; Carriquiry et al., 2020; USDA:APHIS: VS, 2020).

**Conclusion**

In this study, we developed and deployed a new interpretable machine learning methodology capable of benchmarking PRRSV outbreak risk for production systems and individual farms based on biosecurity levels, identifying contributions of key biosecurity aspects on the risk of PRRSV outbreaks in sow farms which can ultimately guide the implementation and enhancement of biosecurity practices. Our findings demonstrate that sharing hauling trailers, increasing numbers and turnover of employees, and a high density of swine in the surrounding area had particularly strong influences on the predicted risk of new PRRSV outbreaks, highlighting the importance of biosecurity practices to mitigate fomite spread, local transmission and improvements in biosecurity training. This study also displayed reasonable agreement in the interpretation of variable contribution in predicting PRRSV outbreak, between the global and local interpretable machine learning models. Importantly, through the local model we developed an individualized approach to assess and guide biosecurity implementation at farms on a case-by-case basis. This novel approach provides each farm with an explanatory view of which biosecurity practices and farm demographics are contributing to their predicted risk of PRRSV outbreak. Furthermore, the *MrIML-biosecurity* methodology described here has the potential to expand the applications of machine learning within biosecurity throughout the agricultural sector to tackle other industry relevant diseases.

**Acknowledgments**
This work was supported by Critical Agricultural Research and Extension 2019-68008-29910 from the USDA National Institute of Food and Agriculture. Funding was also provided by the Swine Health Information Center (SHIC). The authors would like to acknowledge participating companies and veterinarians.

**Authors' contributions**
ALS, GSS, DJH, DCLL, and GM conceived the study. GSS, DJH, OO, DCLL, and GM participated in the design of the study including data collection. ALS conducted data processing, cleaning, developed machine learning functions, R package, vignette and the shiny application with the assistance of GM. ALS, GSS, DJH, OO, BWM, DCLL, and GM wrote and edited the manuscript. All authors discussed the results and critically reviewed the manuscript. DCLL and GM secured funding.

**Conflict of interest**

All authors confirm that there are no conflicts of interest to declare

**Ethical statement**

The authors confirm the ethical policies of the journal, as noted on the journal's author guidelines page. Since this work did not involve animal sampling nor questionnaire data collection by the researchers there was no need for ethics permits.

**Data Availability Statement**

The data that support the findings of this study are not publicly available and are protected by confidential agreements, therefore, are not available

### References

Alarcón, L.V., A. Allepuz, and E. Mateu, 2021: Biosecurity in pig farms: a review. *Porc Health Manag* **7**, DOI: 10.1186/s40813-020-00181-z.

Arruda, A.G., S. Tousignant, J. Sanhueza, C. Vilalta, Z. Poljak, M. Torremorell, C. Alonso, and C.A. Corzo, 2019: Aerosol Detection and Transmission of Porcine Reproductive and Respiratory Syndrome Virus (PRRSV): What Is the Evidence, and What Are the Knowledge Gaps? *Viruses* **11**, 712, DOI: 10.3390/v11080712.

Baekbo, P., and C.S. Kristensen, 2015: PRRS control and eradication plans in Europe. p. 20. In: *Book of Abstracts*. Presented at the International PRRS Congress, Ghent, Belgium.

Barredo Arrieta, A., N. Díaz-Rodríguez, J. Del Ser, A. Bennetot, S. Tabik, A. Barbado, S. Garcia, S. Gil-Lopez, D. Molina, R. Benjamins, R. Chatila, and F. Herrera, 2020: Explainable Artificial Intelligence (XAI): Concepts, taxonomies, opportunities and challenges toward responsible AI. *Information Fusion* **58**, 82–115, DOI: 10.1016/j.inffus.2019.12.012.

Biecek, P., and T. Burzykowski, 2021: Explanatory Model Analysis: Explore, Explain and Examine Predictive Models. New York, NY, USA: Chapman and Hall/CRC Press**.**

Bierk, M.D., S.A. Dee, K.D. Rossow, S. Otake, J.E. Collins, T.W. Molitor, 2001: Transmission of porcine reproductive and respiratory syndrome virus from persistently infected sows to contact controls. *Can J Vet Res* **65**, 261-266.

Black, N.J., L.E. Moraes, and A.G. Arruda, 2021: Association between different types of within-farm worker movements and number of pigs weaned per sow in U.S. Swine farms. *Prev Vet Med* **186**, 105207, DOI: 10.1016/j.prevetmed.2020.105207.

Boehringer Ingelheim, 2018: boehringer-ingelheim.com, Boehringer Ingelheim launches a tool to aid in the prevention of African swine fever [Online] Available at https://www.boehringer-ingelheim.com/press-release/tool-aid-prevention-african-swine-fever (accessed April 10, 2021).

Boessen, C., G. Artz, and L. Schulz, 2018: A baseline study of labor issues and trends in U.S. pork production p. 44. . National Pork Producers Council.

Boser, B.E., I.M. Guyon, and V.N. Vapnik, 1992: A training algorithm for optimal margin classifiers. pp. 144–152. In: *COLT '92: Proceedings of the fifth annual workshop on Computational learning theory*. Presented at the COLT: Annual Workshop on Computational Learning Theory, Pittsburgh, Pennsylvania, United States: ACM Press.

Boulesteix, A.-L., S. Janitza, A. Hapfelmeier, K. Van Steen, and C. Strobl, 2015: Letter to the Editor: On the term 'interaction' and related phrases in the literature on Random Forests. *Brief Bioinform* **16**, 338–345, DOI: 10.1093/bib/bbu012.

Breiman, L., 1996: Stacked regressions. *Mach Learn* **24**, 49–64, DOI: 10.1007/BF00117832.

Breiman, L., 2001: Random forests. *Mach Learn* **45**, 5–32, DOI: 10.1023/A:1010933404324.

Carriquiry, M., A. Elobeid, D. Swenson, and D. Hayes, 2020: Impacts of African Swine Fever in Iowa and the United States. *CARD Working Papers* **[20-WP 600]**, 28.

Carvalho, D.V., E.M. Pereira, and J.S. Cardoso, 2019: Machine Learning Interpretability: A Survey on Methods and Metrics. *Electronics* **8**, 832, DOI: 10.3390/electronics8080832.

Chakraborty, D., H. Başağaoğlu, and J. Winterle, 2021: Interpretable vs. noninterpretable machine learning models for data-driven hydro-climatological process modeling. *Expert Systems with Applications* **170**, DOI: 10.1016/j.eswa.2020.114498.

Chawla, N.V., K.W. Bowyer, L.O. Hall, and W.P. Kegelmeyer, 2002: SMOTE: Synthetic Minority Over-sampling Technique. *jair* **16**, 321–357, DOI: 10.1613/jair.953.

Chicco, D., and G. Jurman, 2020: The advantages of the Matthews correlation coefficient (MCC) over F1 score and accuracy in binary classification evaluation. *BMC Genomics* **21**, DOI: 10.1186/s12864-019-6413-7.

Chicco, D., N. Tötsch, and G. Jurman, 2021: The Matthews correlation coefficient (MCC) is more reliable than balanced accuracy, bookmaker informedness, and markedness in two-class confusion matrix evaluation. *BioData Mining* **14**, DOI: 10.1186/s13040-021-00244-z.

Corzo, C.A., E. Mondaca, S. Wayne, M. Torremorell, S. Dee, P. Davies, and R.B. Morrison, 2010: Control and elimination of porcine reproductive and respiratory syndrome virus. *Virus Res* **154**, 185–192, DOI: 10.1016/j.virusres.2010.08.016.

Dee, S., J. Deen, D. Burns, G. Douthit, and C. Pijoan, 2004: An assessment of sanitation protocols for commercial transport vehicles contaminated with porcine reproductive and respiratory syndrome virus. *Can J Vet Res* **68**, 208–214.

Dee, S., S. Otake, S. Oliveira, and J. Deen, 2009: Evidence of long distance airborne transport of porcine reproductive and respiratory syndrome virus and Mycoplasma hyopneumoniae. *Vet Res* **40**, 39, DOI: 10.1051/vetres/2009022.

Dietterich, T.G., 2000: Ensemble Methods in Machine Learning. pp. 1–15. In: *Multiple Classifier Systems*. Berlin, Heidelberg: Springer.

Eclercy, J., P. Renson, A. Lebret, E. Hirchaud, V. Normand, M. Andraud, F. Paboeuf, Y. Blanchard, N. Rose, and O. Bourry, 2019: A Field Recombinant Strain Derived from Two Type 1 Porcine Reproductive and Respiratory Syndrome Virus (PRRSV-1) Modified Live Vaccines Shows Increased Viremia and Transmission in SPF Pigs. *Viruses* **11**, 296, DOI: 10.3390/v11030296.

Elith, J., J.R. Leathwick, and T. Hastie, 2008: A working guide to boosted regression trees. *Journal of Animal Ecology* **77**, 802–813, DOI: https://doi.org/10.1111/j.1365-2656.2008.01390.x.

Evans, C.M., G.F. Medley, and L.E. Green, 2008: Porcine reproductive and respiratory syndrome virus (PRRSV) in GB pig herds: farm characteristics associated with heterogeneity in seroprevalence. *BMC Veterinary Research* **4**, 48, DOI: 10.1186/1746-6148-4-48.

Ezanno, P., S. Picault, G. Beaunée, X. Bailly, F. Muñoz, R. Duboz, H. Monod, and J.-F. Guégan, 2021: Research perspectives on animal health in the era of artificial intelligence. *Vet Res* **52**, 40, DOI: 10.1186/s13567-021-00902-4.

FAO, OIE - World Organisation for Animal Health, and Weltbank (Eds.), 2010: Good Practices

for Biosecurity in the Pig Sector: Issues and Options in Developing and Transition CountriesRepr. 2010 (July). Rome.

Fauvel, K., V. Masson, É. Fromont, P. Faverdin, and A. Termier, 2019: Towards Sustainable Dairy Management - A Machine Learning Enhanced Method for Estrus Detection. pp. 3051–3059. In: *Proceedings of the 25th ACM SIGKDD International Conference on Knowledge Discovery & Data Mining*. New York, NY, USA: Association for Computing Machinery.

Fountain-Jones, N., C. Kozakiewicz, B. Forester, E. Landguth, S. Carver, M. Charleston, R. Gagne, B. Greenwell, S. Kraberger, D. Trumbo, M. Mayer, N. Clark, and G. Machado, 2021: MrIML: Multi-response interpretable machine learning to map genomic landscapes *Molecular Ecology Resources*, DOI: 10.1111/1755-0998.13495.

Fountain-Jones, N.M., G. Machado, S. Carver, C. Packer, M. Recamonde-Mendoza, and M.E. Craft, 2019: How to make more from exposure data? An integrated machine learning pipeline to predict pathogen exposure. (Andy Fenton, Ed.) *J Anim Ecol* **88**, 1447–1461, DOI: 10.1111/1365-2656.13076.

Friedman, J.H., 2001: Greedy Function Approximation: A Gradient Boosting Machine. *The Annals of Statistics* **29**, 1189–1232.

Friedman, J.H., and B.E. Popescu, 2008: Predictive learning via rule ensembles. *The Annals of Applied Statistics* **2**, 916–954, DOI: 10.1214/07-AOAS148.

Galvis, J.A., J.M. Prada, C.A. Corzo, and G. Machado, 2021: Modeling the transmission and vaccination strategy for porcine reproductive and respiratory syndrome virus. *Transboundary and Emerging Diseases*tbed.14007, DOI: 10.1111/tbed.14007.

Gelaude, P., M. Schlepers, M. Verlinden, M. Laanen, and J. Dewulf, 2014: Biocheck.UGent: A quantitative tool to measure biosecurity at broiler farms and the relationship with technical performances and antimicrobial use. *Poultry Science* **93**, 2740–2751, DOI: 10.3382/ps.2014-04002.

Ghai, B., Q.V. Liao, Y. Zhang, R. Bellamy, and K. Mueller, 2021: Explainable Active Learning (XAL): Toward AI Explanations as Interfaces for Machine Teachers. *Proc. ACM Hum.-Comput. Interact.* **4**, 235:1–235:28, DOI: 10.1145/3432934.

Ghent University, 2021: Home | Biocheck.UGent [Online] Available at https://biocheck.ugent.be/en (accessed January 25, 2021).

Goldstein, A., A. Kapelner, J. Bleich, and E. Pitkin, 2015: Peeking Inside the Black Box: Visualizing Statistical Learning With Plots of Individual Conditional Expectation. *Journal of Computational and Graphical Statistics* **24**, 44–65, DOI: 10.1080/10618600.2014.907095.

Greenwell, B.M., 2017: pdp: An R Package for Constructing Partial Dependence Plots. **9**, 16.

Holtkamp, D, J.B. Kliebenstein, and E.J. Neumann, 2013: Assessment of the economic impact of porcine reproductive and respiratory syndrome virus on United States pork producers. *Journal of Swine Health and Production* **21**, 13.

Holtkamp, D.J., D.D. Polson, M. Torremorell, 2011: Terminology for classifying swine herds by porcine reproductive and respiratory syndrome virus status. *JSHAP* **19**, 44-56.

Holtkamp, Derald, H. Lin, W. Chong, and A.M. O'Connor, 2012: Identifying questions in the American Association of Swine Veterinarian's PRRS risk assessment survey that are important for retrospectively classifying swine herds according to whether they reported clinical PRRS outbreaks in the previous 3 years. *Preventive Veterinary Medicine* **106**, 42–52, DOI: 10.1016/j.prevetmed.2012.03.003.

Hornung, R., and A.-L. Boulesteix, 2021: Interaction Forests: Identifying and exploiting interpretable quantitative and qualitative interaction effects p. 33. . University of Munich.

Jara, M., D.A. Rasmussen, C.A. Corzo, and G. Machado, 2020: Porcine reproductive and respiratory syndrome virus dissemination across pig production systems in the United States. *bioRxiv*2020.04.09.034181, DOI: 10.1101/2020.04.09.034181.

Jurado, C., L. Mur, M.S. Pérez Aguirreburualde, E. Cadenas-Fernández, B. Martínez-López, J.M. Sánchez-Vizcaíno, and A. Perez, 2019: Risk of African swine fever virus introduction into the United States through smuggling of pork in air passenger luggage. *Scientific Reports* **9**, 14423, DOI: 10.1038/s41598-019-50403-w.

Kopitar, L., L. Cilar, P. Kocbek, and G. Stiglic, 2019: Local vs. Global Interpretability of Machine Learning Models in Type 2 Diabetes Mellitus Screening. pp. 108–119. In: Marcos, Mar, Jose M. Juarez, Richard Lenz, Grzegorz J. Nalepa, Slawomir Nowaczyk, Mor Peleg, Jerzy Stefanowski, and Gregor Stiglic (eds), *Artificial Intelligence in Medicine: Knowledge Representation and Transparent and Explainable Systems*. Cham: Springer International Publishing.

Krishna, V.D., Y. Kim, M. Yang, F. Vannucci, T. Molitor, M. Torremorell, and M.C.-J. Cheeran, 2020: Immune responses to porcine epidemic diarrhea virus (PEDV) in swine and protection against subsequent infection. *PLOS ONE* **15**, e0231723, DOI: 10.1371/journal.pone.0231723.

Kruse, A.B., L.R. Nielsen, and L. Alban, 2020: Herd typologies based on multivariate analysis of biosecurity, productivity, antimicrobial and vaccine use data from Danish sow herds. *Preventive Veterinary Medicine* **181**, 104487, DOI: 10.1016/j.prevetmed.2018.06.008.

Kuhn, M., and H. Wickham, 2021: Discretize Numeric Variables — discretize [Online] Available at https://recipes.tidymodels.org/reference/discretize.html (accessed April 13, 2021).

Lambert, M.-È., J. Arsenault, Z. Poljak, and S. D'Allaire, 2012: Epidemiological investigations in regard to porcine reproductive and respiratory syndrome (PRRS) in Quebec, Canada. Part 2: Prevalence and risk factors in breeding sites. *Preventive Veterinary Medicine* **104**, 84–93, DOI: 10.1016/j.prevetmed.2011.11.002.

Lee, C., 2015: Porcine epidemic diarrhea virus: An emerging and re-emerging epizootic swine virus. *Virology Journal* **12**, 193, DOI: 10.1186/s12985-015-0421-2.

Lee, J., L.L. Schulz, and G.T. Tonsor, 2021: Swine producer willingness to pay for Tier 1 disease risk mitigation under multifaceted ambiguity. *Agribusiness* **n/a**, DOI: https://doi.org/10.1002/agr.21694.

Li, J., D. Lin, Y. Wang, G. Xu, and C. Ding, 2021: Towards a Reliable Evaluation of Local Interpretation Methods. *Applied Sciences* **11**, 2732, DOI: 10.3390/app11062732.

Liu, P., Y. Bai, X. Jiang, L. Zhou, S. Yuan, H. Yao, H. Yang, and Z. Sun, 2018: High reversion potential of a cell-adapted vaccine candidate against highly pathogenic porcine reproductive and respiratory syndrome. *Veterinary Microbiology* **227**, 133–142, DOI: 10.1016/j.vetmic.2018.10.004.

Ljubobratovic, D., Z. Guoxiang, M. Brkic Bakaric, T. Jemric, and M. Matetic, 2020: Predicting Peach Fruit Ripeness Using Explainable Machine Learning. Vol. 1, pp. 0717–0723. In: Katalinic, Branko (ed), *DAAAM Proceedings*. DAAAM International Vienna.

Lucas, T.C.D., 2020: A translucent box: interpretable machine learning in ecology. *Ecol Monogr* **90**, DOI: 10.1002/ecm.1422.

Lunardon, N., G. Menardi, and N. Torelli, 2014: ROSE: a Package for Binary Imbalanced Learning. *The R Journal* **6**, 79, DOI: 10.32614/RJ-2014-008.

Machado, G., 2021: mrIML: Multivariate–multi-response–interpretable machine learning, Multi Response Interpretable Machine Learning [Online] Available at https://nfj1380.github.io/mrIML/ (accessed February 4, 2021). github.

Machado, G., M.R. Mendoza, and L.G. Corbellini, 2015: What variables are important in predicting bovine viral diarrhea virus? A random forest approach. *Veterinary Research* **46**, 85, DOI: 10.1186/s13567-015-0219-7.

Machado, G., C. Vilalta, M. Recamonde-Mendoza, C. Corzo, M. Torremorell, A. Perez, and K. VanderWaal, 2019: Identifying outbreaks of Porcine Epidemic Diarrhea virus through animal movements and spatial neighborhoods. *Sci Rep* **9**, 457, DOI: 10.1038/s41598-018-36934-8.

Mandrekar, J.N., 2010: Receiver Operating Characteristic Curve in Diagnostic Test Assessment. *Journal of Thoracic Oncology* **5**, 1315–1316, DOI: 10.1097/JTO.0b013e3181ec173d.

Merrill, S.C., C.J. Koliba, S.M. Moegenburg, A. Zia, J. Parker, T. Sellnow, S. Wiltshire, G. Bucini, C. Danehy, J.M. Smith, 2019: Decision-making in livestock biosecurity practices amidst environmental and social uncertainty: Evidence from an experimental game. *PLOS ONE* **14**, e0214500, DOI: 10.1371/journal.pone.0214500

Molnar, C., 2021: Interpretable Machine Learning.

Montaner-Tarbes, S., H.A. del Portillo, M. Montoya, and L. Fraile, 2019: Key Gaps in the Knowledge of the Porcine Respiratory Reproductive Syndrome Virus (PRRSV). *Frontiers in Veterinary Science* **6**, 38, DOI: 10.3389/fvets.2019.00038.

Mortensen, S., H. Stryhn, R. Søgaard, A. Boklund, K.D.C. Stärk, J. Christensen, and P. Willeberg, 2002: Risk factors for infection of sow herds with porcine reproductive and respiratory syndrome (PRRS) virus. *Preventive Veterinary Medicine* **53**, 83–101, DOI: 10.1016/S0167-5877(01)00260-4.

Nathues, H., P. Alarcon, J. Rushton, R. Jolie, K. Fiebig, M. Jimenez, V. Geurts, and C. Nathues, 2018: Modelling the economic efficiency of using different strategies to control Porcine Reproductive & Respiratory Syndrome at herd level. *Preventive Veterinary Medicine* **152**, 89–102, DOI: 10.1016/j.prevetmed.2018.02.005.

Neethirajan, S., 2020: The role of sensors, big data and machine learning in modern animal farming. *Sensing and Bio-Sensing Research* **29**, 100367, DOI: 10.1016/j.sbsr.2020.100367.

Neumann, E.J., J.B. Kliebenstein, C.D. Johnson, J.W. Mabry, E.J. Bush, A.H. Seitzinger, A.L. Green, and J.J. Zimmerman, 2005: Assessment of the economic impact of porcine reproductive and respiratory syndrome on swine production in the United States. *Journal of the American Veterinary Medical Association* **227**, 385–392, DOI: 10.2460/javma.2005.227.385.

Oh, S., 2019: Feature Interaction in Terms of Prediction Performance. *Applied Sciences* **9**, 5191, DOI: 10.3390/app9235191.

Otake, S., S.A. Dee, R.D. Moon, K.D. Rossow, C. Trincado, and C. Pijoan, 2004: Studies on the carriage and transmission of porcine reproductive and respiratory syndrome virus by individual houseflies (Musca domestica). *Vet Rec* **154**, 80–85, DOI: 10.1136/vr.154.3.80.

Otake, Satoshi, S. Dee, C. Corzo, S. Oliveira, and J. Deen, 2010: Long-distance airborne transport of infectious PRRSV and Mycoplasma hyopneumoniae from a swine population infected with multiple viral variants. *Vet Microbiol* **145**, 198–208, DOI: 10.1016/j.vetmic.2010.03.028.

Otake, Satoshi, S.A. Dee, R.D. Moon, K.D. Rossow, C. Trincado, M. Farnham, and C. Pijoan, 2003: Survival of porcine reproductive and respiratory syndrome virus in houseflies. *Can J Vet Res* **67**, 198–203.

Otake, Satoshi, S.A. Dee, K.D. Rossow, J. Deen, H.S. Joo, T.W. Molitor, and C. Pijoan, 2002: Transmission of porcine reproductive and respiratory syndrome virus by fomites (boots and coveralls). *Journal of Swine Health and Production* **10**, 59–65.

Pileri, E., and E. Mateu, 2016: Review on the transmission porcine reproductive and respiratory syndrome virus between pigs and farms and impact on vaccination. *Vet Res* **47**, 108, DOI: 10.1186/s13567-016-0391-4.

Pitkin, A., J. Deen, and S. Dee, 2009: Further assessment of fomites and personnel as vehicles for the mechanical transport and transmission of porcine reproductive and respiratory syndrome virus. *Can J Vet Res* **73**, 298–302.

Polikar, R., 2012: Ensemble Learning, In: Zhang, Cha, and Yunqian Ma (eds), Ensemble Machine Learning. Boston, MA: Springer US.

Pudenz, C.C., L.L. Schulz, and G.T. Tonsor, 2019: Adoption of Secure Pork Supply Plan Biosecurity by U.S. Swine Producers. *Front Vet Sci* **6**, DOI: 10.3389/fvets.2019.00146.

Rabinovich, J.E., A.A. Costa, I.J. Muñoz, P.E. Schilman, and N.M. Fountain-Jones, 2021: Machine-learning model led design to experimentally test species thermal limits: The case of kissing bugs (Triatominae). *PLOS Neglected Tropical Diseases* **15**, e0008822, DOI: 10.1371/journal.pntd.0008822.

Rabinowitz, P., H. Fowler, L.O. Odofin, C. Messinger, J. Sparer, and S. Vegso, 2013: Swine worker awareness and behavior regarding prevention of zoonotic influenza transmission. *J Agromedicine* **18**, 304–311, DOI: 10.1080/1059924X.2013.826603.

Racicot, M., D. Venne, A. Durivage, and J.-P. Vaillancourt, 2012: Evaluation of the relationship between personality traits, experience, education and biosecurity compliance on poultry farms in Québec, Canada. *Prev Vet Med* **103**, 201–207, DOI: 10.1016/j.prevetmed.2011.08.011.

Ramirez, A., and P. Zaabel, 2012: Swine Biological Risk Management. PhD thesis, Iowa State University.

Rathkjen, P.H., and J. Dall, 2017: Control and eradication of porcine reproductive and respiratory syndrome virus type 2 using a modified-live type 2 vaccine in combination with a load, close, homogenise model: an area elimination study. *Acta Vet Scand* **59**, 4, DOI: 10.1186/s13028-016-0270-z.

Renken, C., C. Nathues, H. Swam, K. Fiebig, C. Weiss, M. Eddicks, M. Ritzmann, and H. Nathues, 2021: Application of an economic calculator to determine the cost of porcine reproductive and respiratory syndrome at farm-level in 21 pig herds in Germany. *Porcine Health Management* **7**, 3, DOI: 10.1186/s40813-020-00183-x.

Ribeiro, M.T., S. Singh, and C. Guestrin, 2016: Nothing Else Matters: Model-Agnostic Explanations By Identifying Prediction Invariance. *arXiv:1611.05817 [cs, stat]*.

Rodrigues da Costa, M., J. Gasa, J.A. Calderón Díaz, M. Postma, J. Dewulf, G. McCutcheon, and E.G. Manzanilla, 2019: Using the Biocheck.UGent^TM scoring tool in Irish farrow-to-finish pig farms: assessing biosecurity and its relation to productive performance. *Porc Health Manag* **5**, 4, DOI: 10.1186/s40813-018-0113-6.

Romagosa, A., 2017: Applied review on evidence-based biosecurity. pp. 5–11. . Presented at the 48th AASV Annual Meeting, Denver, Colorado.

Sasaki, Y., A. Furutani, T. Furuichi, T. Hayakawa, S. Ishizeki, R. Kano, F. Koike, M. Miyashita, Y. Mizukami, Y. Watanabe, and S. Otake, 2020: Development of a biosecurity assessment tool and the assessment of biosecurity levels by this tool on Japanese commercial swine farms. *Preventive Veterinary Medicine* **175**, 104848, DOI: 10.1016/j.prevetmed.2019.104848.

Silva, G.S., L.G. Corbellini, D.L.C. Linhares, K.L. Baker, and D.J. Holtkamp, 2018: Development and validation of a scoring system to assess the relative vulnerability of swine breeding herds to the introduction of PRRS virus. *Preventive Veterinary Medicine* **160**, 116–122, DOI: 10.1016/j.prevetmed.2018.10.004.

Silva, G.S., G. Machado, K.L. Baker, D.J. Holtkamp, and D.C.L. Linhares, 2019: Machine-learning algorithms to identify key biosecurity practices and factors associated with breeding herds reporting PRRS outbreak. *Prev Vet Med* **171**, 104749, DOI: 10.1016/j.prevetmed.2019.104749.

Song, D., H. Moon, and B. Kang, 2015: Porcine epidemic diarrhea: a review of current epidemiology and available vaccines. *Clinical and Experimental Vaccine Research* **4**, 166–176, DOI: 10.7774/cevr.2015.4.2.166.

Staniak, M., and P. Biecek, 2019: Explanations of Model Predictions with live and breakDown Packages. *The R Journal* **10**, 395, DOI: 10.32614/RJ-2018-072.

Strobl, C., J. Malley, and G. Tutz, 2009: An Introduction to Recursive Partitioning: Rationale, Application and Characteristics of Classification and Regression Trees, Bagging and

Random Forests. *Psychol Methods* **14**, 323–348, DOI: 10.1037/a0016973.

Tousignant, S.J.P., A. Perez, and R. Morrison, 2015: Comparison between the 2013-2014 and 2009-2012 annual porcine reproductive and  respiratory syndrome virus epidemics in a cohort of sow herds in the United States. *Can Vet J* **56**, 1087–1089.

USDA:APHIS:VS, 2020: USDA APHIS ASF Response Plan: The Red Book. .

Valdes-Donoso, P., L.S. Jarvis, D. Wright, J. Alvarez, and A.M. Perez, 2016: Measuring Progress on the Control of Porcine Reproductive and Respiratory Syndrome (PRRS) at a Regional Level: The Minnesota N212 Regional Control Project (Rcp) as a Working Example. (Frederick C. Leung, Ed.)*PLoS ONE* **11**, e0149498, DOI: 10.1371/journal.pone.0149498.

Velasova, M., P. Alarcon, S. Williamson, and B. Wieland, 2012: Risk factors for porcine reproductive and respiratory syndrome virus infection and resulting challenges for effective disease surveillance. *BMC Veterinary Research* **8**, 184, DOI: 10.1186/1746-6148-8-184.

Vilata, C., J. Sanhueza, M. Kikuti, C. Corzo, 2019: pig333.com, Porcine Reproductive and Respiratory Syndrome break recurrence [Online] Available at https://www.pig333.com/articles/porcine-reproductive-and-respiratory-syndrome-break-recurrence_14543/

Wright, M.N., A. Ziegler, and I.R. König, 2016: Do little interactions get lost in dark random forests? *BMC Bioinformatics* **17**, 145, DOI: 10.1186/s12859-016-0995-8.

Yang, G., Q. Ye, and J. Xia, 2021: Unbox the Black-box for the Medical Explainable AI via Multimodal and Multi-centre Data Fusion: A Mini-Review, Two Showcases and Beyond. *arXiv:2102.01998 [cs, math]*.

Zimmerman, J.J., S.A. Dee, D.J. Holtkamp, M.P. Murtaugh, T. Stadejek, G.W. Stevenson, M. Torremorell, H. Yang, and J. Zhang, 2019: Porcine Reproductive and Respiratory Syndrome Viruses (Porcine Arteriviruses), pp. 685–708. In: Diseases of Swine. John Wiley & Sons, Ltd.

**Figures**

**Figure 1. Flow chart of the implemented machine learning pipeline and related outputs.** (a) Data regarding biosecurity practices and farm demographics and the number of PRRSV outbreaks in the past five years were collected via a 42 question survey (Silva et al., 2019). (b) Balancing techniques were implemented due to the uneven distribution of outcomes and followed by data partitioning to create test and training data. (c) Individual machine learning models and an ensemble model were trained, employing cross-validation techniques to reduce overfitting. Using performance metrics, the most appropriate model was selected and used to produce all results. (d) Global benchmarking consisted of predicted risk benchmarking among (left panel) and within systems, and model interpretation via variable importance (right panel), partial dependence profiles, and feature interaction. (e) Local benchmarking consisted of the interpretation and benchmarking of biosecurity practice and farm demographic contribution to the predicted PRRSV outbreak risk at individual farms, shown by the *phi* value, presented as aggregated summary box plots (left panel) and individualized waterfall plots (right panel).

**Figure 2. Cumulative importance of biosecurity practices and farm demographics.** The y-axis is (top to bottom) in descending order of importance. Cumulative importance values calculated through the Gini Index are represented by the x-axis. Higher cumulative importance values indicate a higher order of importance in relation to PRRSV outbreaks.

**Figure 3. Distribution of farms reporting PRRSV outbreaks (none versus at least one) by production system in relation to the risk of a new outbreak.** The x-axis represents the production system identification and the y-axis represents the predicted PRRSV outbreak risk produced by the machine learning model. Farms below the dashed line are considered low risk, between the dashed line and the solid line are considered medium risk, and above the solid line are considered high risk.

**Figure 4. Distribution of observed biosecurity practice and farm demographic values by PRRSV outbreak risk category.** The predicted PRRSV outbreak risk values were categorized by a *tidymodels* discretization method to create categories for low, medium, and high predicted outbreak risk (Kuhn and Wickham, 2021). (A) distribution of farms raising genetic breeding replacements; (B) distribution of farm distance from the public road; (C) distribution of the number of breeding females in the surrounding three-miles; and (D) distribution of annual turnover of employees shown via a proportion (0% to 100%).

**Figure 5. The rank of biosecurity practices produced by local model agnostic interpretation methods.** The boxplots represent a summary of the breakDown values generated for each biosecurity practice per farm, ordered by mean *phi* value. The current status of farms is presented as having reported at least one outbreak (red) or having reported no outbreaks (gray). The axis

represents the contribution of the practice to PRRSV outbreak prediction. Values above zero are contributing to an increased risk prediction and values below zero are contributing to a decreased risk prediction.

**Supplementary Material**

**Interpretable machine learning applied to on-farm biosecurity and porcine reproductive and respiratory syndrome virus**

**Running title: Application of machine learning to pig farm biosecurity**

Abagael L. Sykes[1], Gustavo S. Silva[2], Derald J. Holtkamp[2], Broc W. Mauch[2], Onyekachukwu Osemeke[2],  Daniel C.L. Linhares[2], Gustavo Machado[1]*

[1] Department of Population Health and Pathobiology, College of Veterinary Medicine, Raleigh, North Carolina, USA.
[2] Veterinary Diagnostic and Production Animal Medicine Department, College of Veterinary Medicine, Iowa State University, Ames, Iowa, USA.

**\*Corresponding author:** gmachad@ncsu.edu

| Biosecurity practice or farm demographic | Mean | Standard deviation |
|---|---|---|
| Number of swine premises in a 3-mile radius | 2.58 | 3.92 |
| Capacity of breeding females in 3-mile radius | 2647.89 | 4903.85 |
| Capacity of nursery pigs in 3-mile radius | 912.23 | 3767.89 |
| Capacity of finishing pigs in 3-mile radius | 862.59 | 2402.57 |
| Capacity of wean-to-finish pigs in 3-mile radius | 2986.30 | 8347.08 |
| Capacity of gilts in 3-mile radius | 746.12 | 1912.35 |
| Capacity of boars in3-mile radius | 4.97 | 26.69 |
| Capacity of show pigs 3-mile radius | 0.43 | 1.41 |
| Yards to the nearest road | 321.40 | 394.89 |
| Number of vehicle entrances | 1.45 | 0.67 |
| Number of other premises with same semen delivery driver/ vehicle | 8.69 | 3.65 |
| Number of breeding replacement deliveries per year | 16.69 | 16.01 |
| Number of other premises with same haul trailer for breeding replacement | 5.58 | 5.57 |
| Number of cull breeding hauls per month | 2.94 | 2.14 |
| Number of other premises with same haul trailer for cull animals | 3.31 | 2.70 |
| Number of weaned hauls of per month | 7.99 | 5.00 |
| Number of other premises with same haul trailer for wean animals | 1.46 | 2.46 |
| Number of dead animal removals per month | 7.35 | 8.77 |
| Number of feed/feed-ingredients deliveries per month | 14.06 | 10.09 |
| Monthly propane/fuel deliveries | 0.99 | 0.78 |
| Number of garbage collected per month | 3.13 | 1.33 |

**Table S1. Descriptive statistics of numerical biosecurity practices and farm demographics**

Mean and standard deviation values of numeric biosecurity practices and farm demographics collected through the biosecurity survey and included as predictor variables within the machine learning algorithm.

| Biosecurity practice or farm demographic | Mean | Standard deviation |
|---|---|---|
| *Number of on-farm employees* | 12.75 | 6.05 |
| *Number of daily exit/re-entries into barns by employees* | 1.54 | 1.33 |
| *Annual turnover of on-site employees* | 0.42 | 0.25 |
| *Number of repair/maintenance visits per year* | 49.13 | 49.63 |
| *Number of visits from veterinarian/manager per month* | 7.07 | 5.62 |
| *Length of required downtime veterinarians/ managers* | 33.39 | 22.90 |
| *Frequency of grass mowing per month* | 3.51 | 1.09 |
| *Frequency of snow removal per year* | 13.60 | 9.80 |
| *Number of times pork/food products enter the site per month* | 26.68 | 8.74 |
| *Number of manure or effluent removals per month* | 1.63 | 0.82 |
| *Length of required downtime for manure removal personnel* | 22.34 | 23.31 |
| *Length of required downtime for manure removal equipment* | 25.53 | 28.42 |
| *Number of PRRSV outbreaks in last 5 years* | 1.27 | 1.32 |

**Table S2. Descriptive statistics of numerical biosecurity practices and farm demographics (continued)**

Mean and standard deviation values of numeric biosecurity practices and farm demographics collected through the biosecurity survey and included as predictor variables within the machine learning algorithm

| Biosecurity practice or farm demographic | Category | Frequency |
|---|---|---|
| *PRRSV Status* | *Positive* | 89 |
| | *Negative* | 50 |
| *Primarily breeding genetic replacements?* | *Yes* | 23 |
| | *No* | 116 |
| *Are the surrounding farms in the same production system?* | *Yes* | 60 |
| | *No* | 79 |
| *Is the disease status of the surrounding farms known?* | *Yes* | 100 |
| | *No* | 39 |
| *Topography* | *Flat or gentle rolling hills* | 137 |
| | *Steep hills or mountains* | 2 |
| *Is the surrounding area forested?* | *Yes* | 41 |
| | *No* | 98 |
| *Are PRRSV positive animals hauled on cull trailer?* | *Yes* | 81 |
| | *No* | 58 |
| *Are PRRSV positive animals hauled on wean trailer?* | *Yes* | 36 |
| | *No* | 103 |
| *Onsite dead disposal method used* | *None* | 29 |
| | *Compost* | 105 |
| | *Incineration* | 5 |
| *Is the air filtered?* | *Never* | 112 |
| | *Seasonally* | 8 |
| | *Year-round* | 19 |

**Table S3. Descriptive statistics of categorical biosecurity practices and farm demographics**

Frequency values for the categorical biosecurity practices and farm demographics collected through the biosecurity survey and included as predictor variables within the machine learning algorithm.

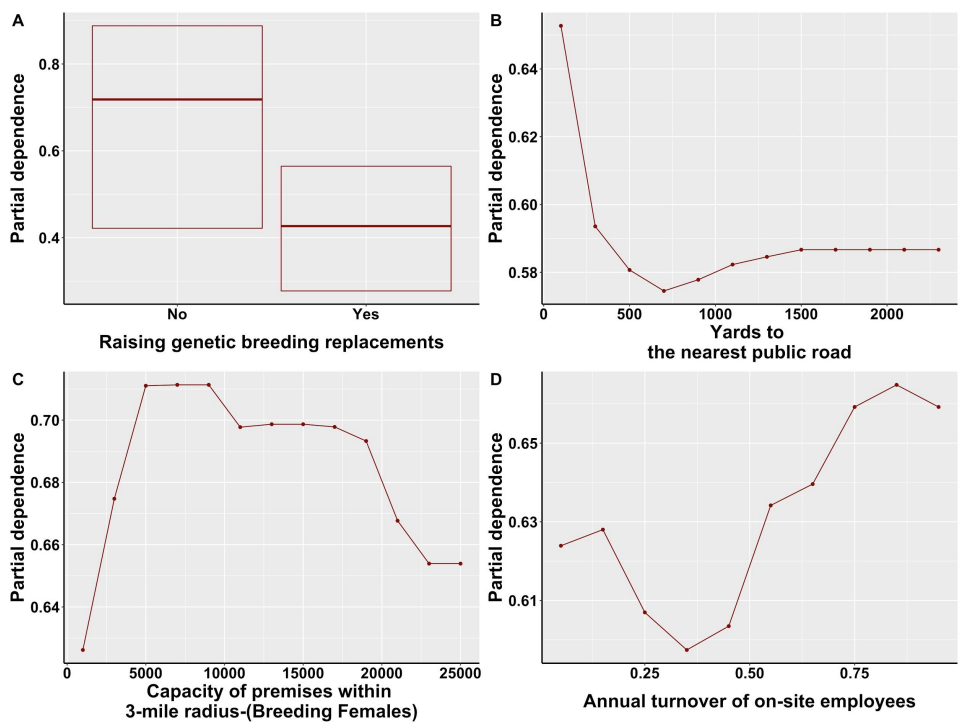

**Figure S1. Partial dependence plots showing the marginal effect of the top four important biosecurity practices and farm demographics on the model.** The following biosecurity practices and farm demographics are presented here: (A) raising genetic replacements; (B) yards to the nearest public road; (C) capacity of breeding females in the surrounding three-miles; and (D) annual turnover of on-siteon site employees. Higher partial dependence values, represented by the y-axis, indicate a large marginal effect of practice or demographic. The x-axis represents the observed values of the practice or demographic.

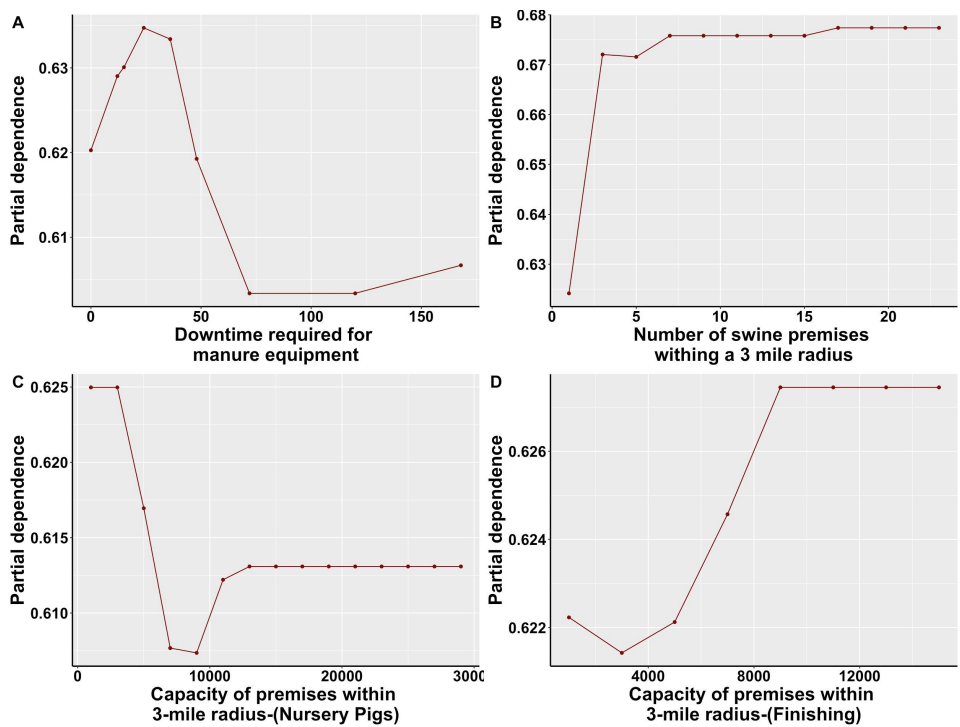

**Figure S2. Partial dependence plots showing the marginal effect of biosecurity practices and farm demographics on the model.** The following biosecurity practices and farm demographics are presented here: (A) downtime required for manure equipment; (B) number of swine farms in a three-mile radius; (C) capacity of nursery pigs in a three-mile radius; and (D) capacity of finishing pigs in a three-mile radius. Higher partial dependence values, represented by the y-axis, indicate a large marginal effect of practice or demographic. The x-axis represents the observed values of the practice or demographic.

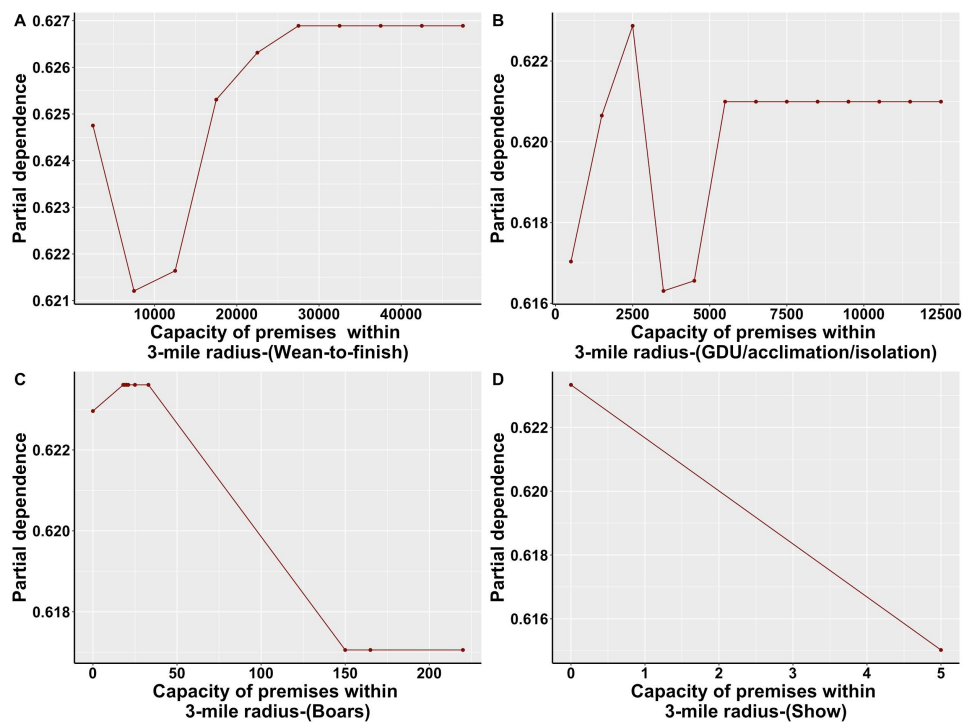

**Figure S3. Partial dependence plots showing the marginal effect of biosecurity practices and farm demographics on the model.** The following biosecurity practices and farm demographics are presented here: (A) capacity of wean-to-finishers in a three-mile radius; (B) capacity of gilts in a three-mile radius; (C) capacity of boars in a three-mile radius; and (D) capacity of show pigs in a three-mile radius. Higher partial dependence values, represented by the y-axis, indicate a large marginal effect of practice or demographic. The x-axis represents the observed values of the practice or demographic.

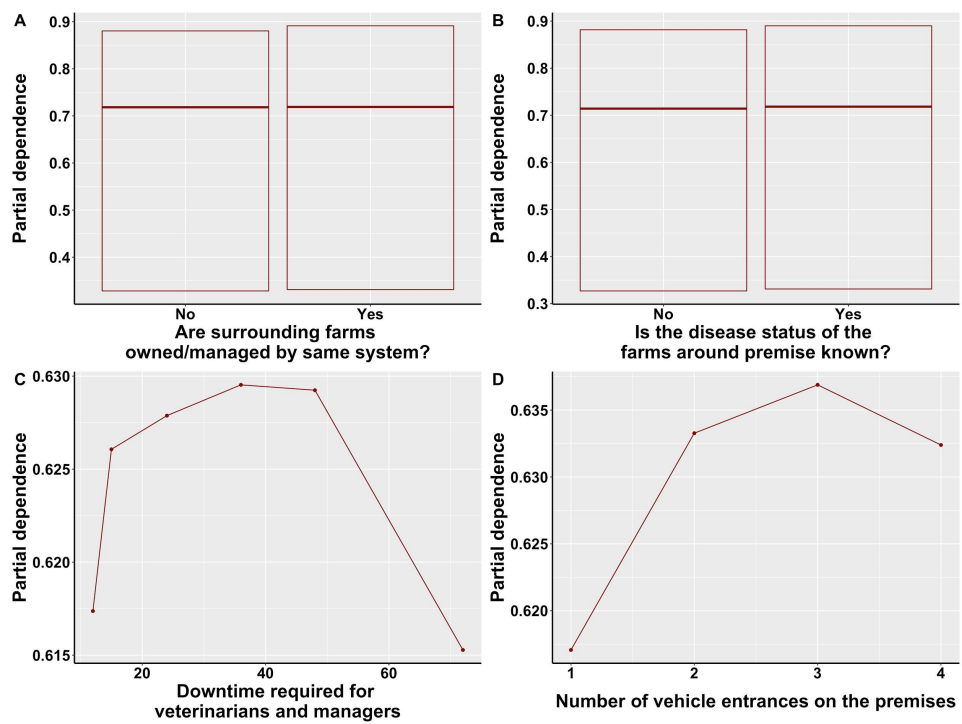

**Figure S4. Partial dependence plots showing the marginal effect of biosecurity practices and farm demographics on the model.** The following biosecurity practices and farm demographics are presented here: (A) whether the surrounding farms are owned or associated with the same system; (B) whether the disease status of the surrounding farms is known; (C) downtime required for veterinarians and managers; and (D) number of vehicle entrances on the premises. Higher partial dependence values, represented by the y-axis, indicate a large marginal effect of practice or demographic. The x-axis represents the observed values of the practice or demographic.

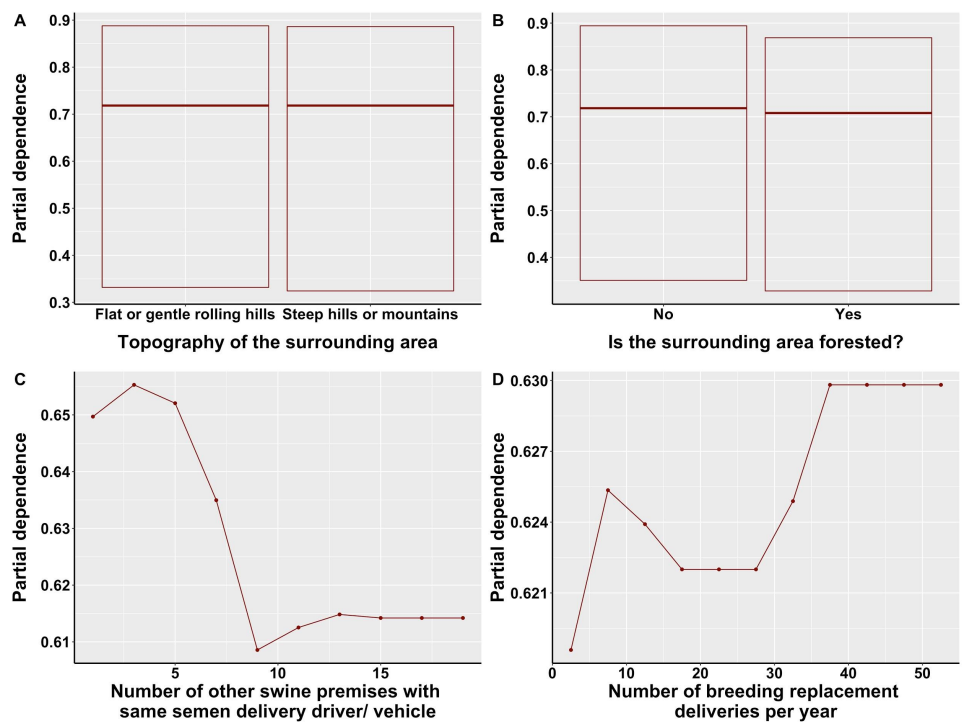

**Figure S5. Partial dependence plots showing the marginal effect of biosecurity practices and farm demographics on the model.** The following biosecurity practices and farm demographics are presented here: (A) topography of the surrounding area; (B) whether the surrounding area is forested; (C) number of other swine premises with the same semen delivery/vehicle; and (D) number of breeding replacement deliveries per year. Higher partial dependence values, represented by the y-axis, indicate a large marginal effect of practice or demographic. The x-axis represents the observed values of the practice or demographic.

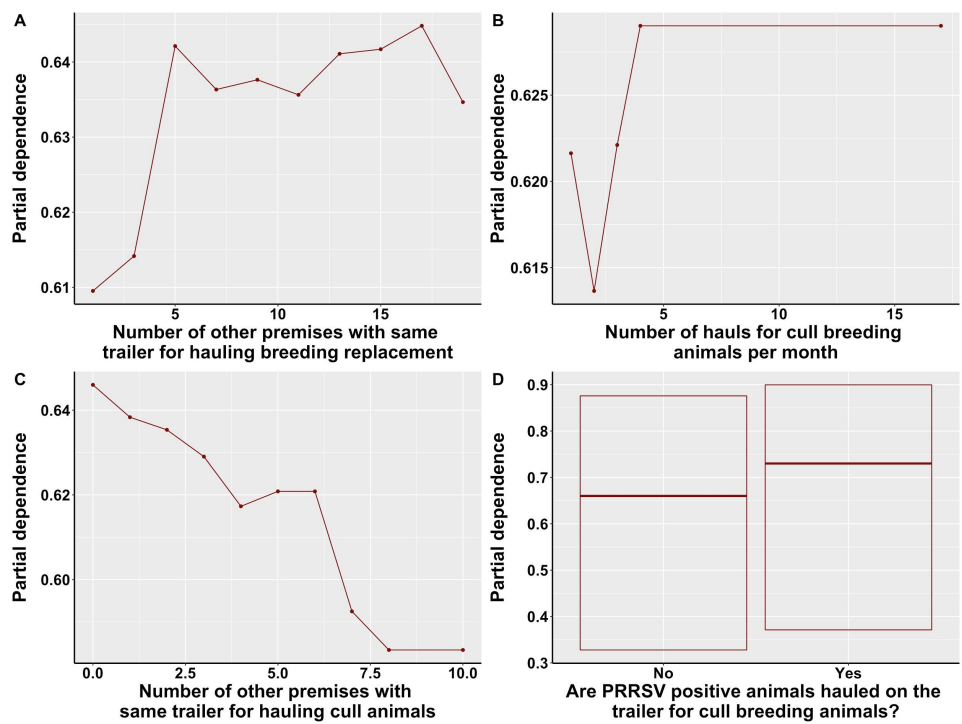

**Figure S6. Partial dependence plots showing the marginal effect of biosecurity practices and farm demographics on the model.** The following biosecurity practices and farm demographics are presented here: (A) number of other premises with the same trailer for hauling breeding replacements; (B) number of hauls for cull breeding animals per month; (C) number of other premises with the same trailer for hauling cull animals; and (D) whether PRRSV positive animals are hauled on the trailer used for cull breeding animals. Higher partial dependence values, represented by the y-axis, indicate a large marginal effect of practice or demographic. The x-axis represents the observed values of the practice or demographic.

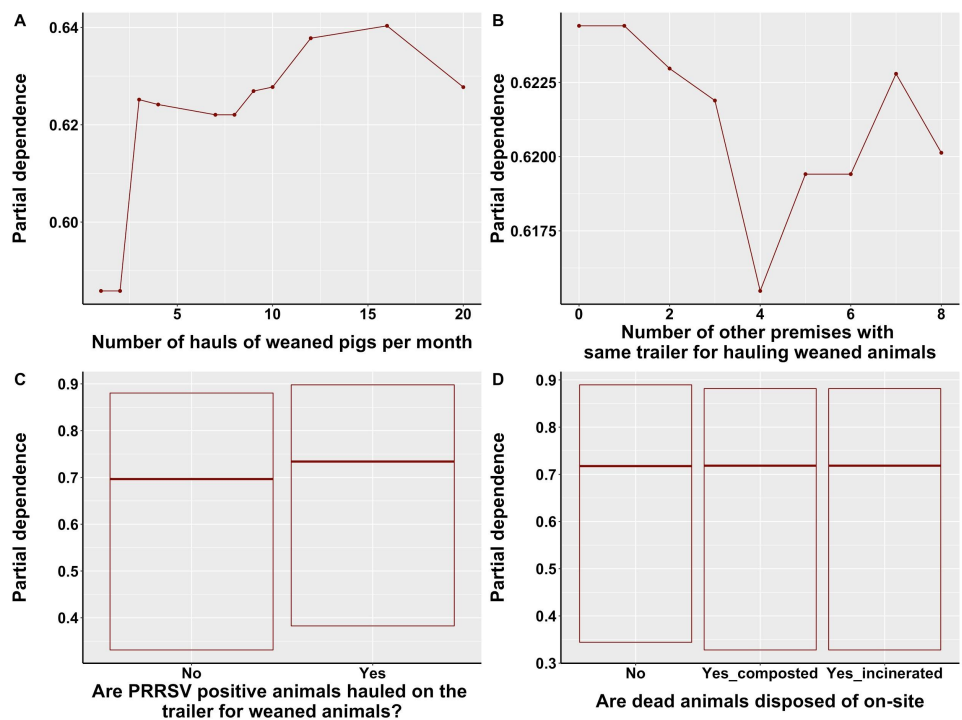

**Figure S7. Partial dependence plots showing the marginal effect of biosecurity practices and farm demographics on the model.** The following biosecurity practices and farm demographics are presented here: (A) number of hauls of weaned pigs per month; (B) number of other premises with same trailer for hauling weaned animals; (C) whether PRRSV positive animals are hauled on the trailed used for weaned animals; and (D) whether dead animals are disposed of on site and which method is used. Higher partial dependence values, represented by the y-axis, indicate a large marginal effect of practice or demographic. The x-axis represents the observed values of the practice or demographic.

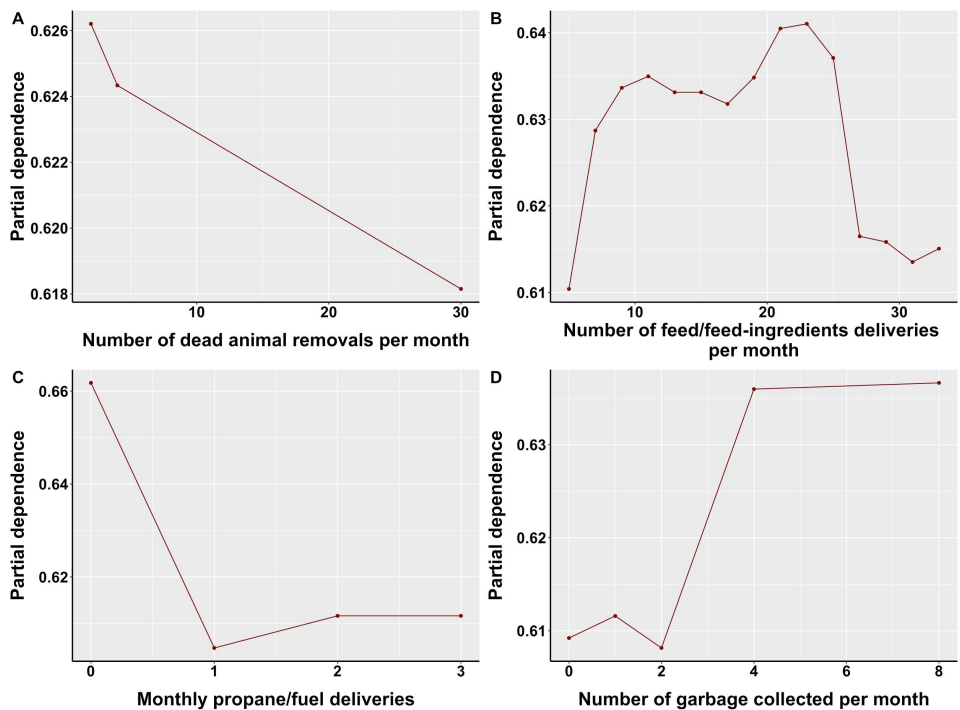

**Figure S8. Partial dependence plots showing the marginal effect of biosecurity practices and farm demographics on the model.** The following biosecurity practices and farm demographics are presented here: (A) number of dead animal removals per month; (B) number of feed/feed-ingredient deliveries per month ; (C) frequency of fuel/propane deliveries per month; and (D) number of garbage collections per month. Higher partial dependence values, represented by the y-axis, indicate a large marginal effect of practice or demographic. The x-axis represents the observed values of the practice or demographic.

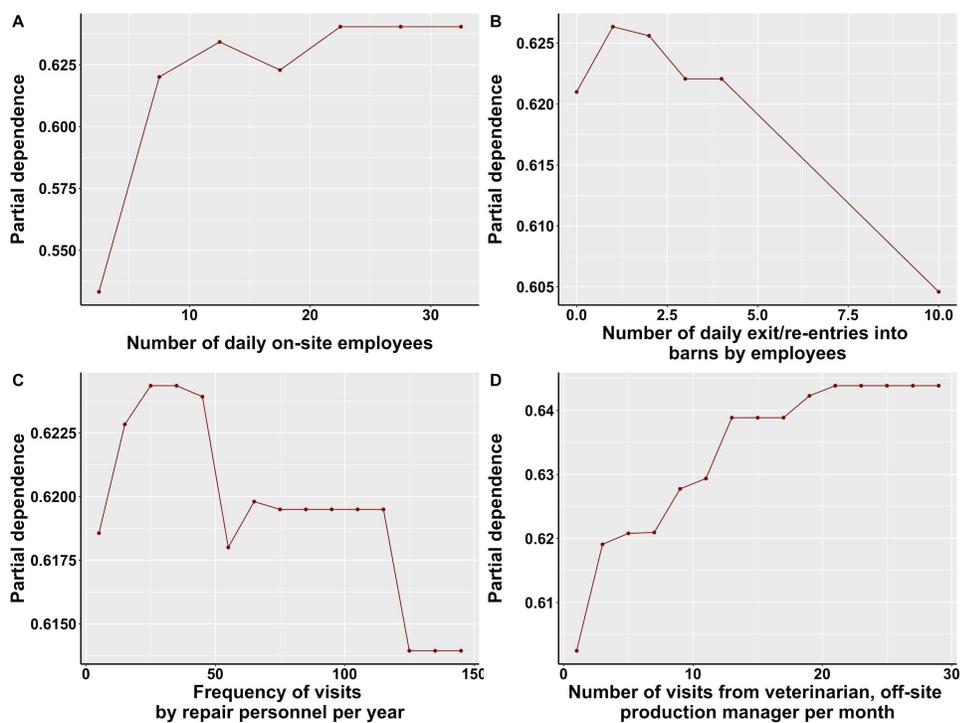

**Figure S9. Partial dependence plots showing the marginal effect of biosecurity practices and farm demographics on the model.** The following biosecurity practices and farm demographics are presented here: (A) number of daily on-site employees; (B) number of exit and reentries unto the barns by employees; (C) number of visits from repair personnel per year; and (D) number of visits from veterinarians and production managers per month. Higher partial dependence values, represented by the y-axis, indicate a large marginal effect of practice or demographic. The x-axis represents the observed values of the practice or demographic.

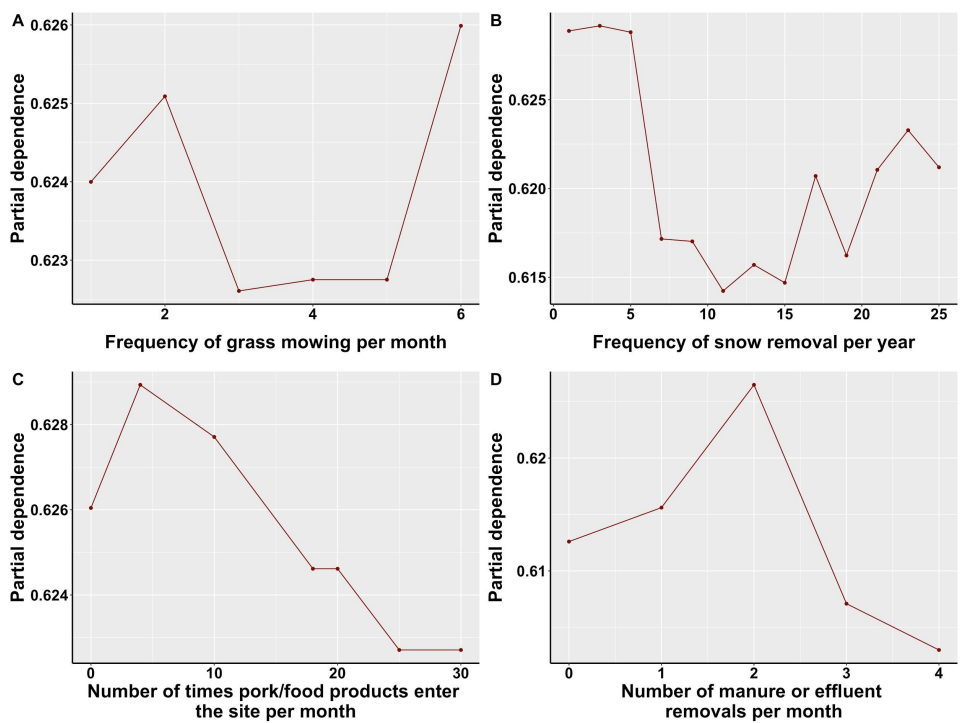

**Figure S10. Partial dependence plots showing the marginal effect of biosecurity practices and farm demographics on the model.** The following biosecurity practices and farm demographics are presented here: (A) frequency of grass mowing per month; (B) frequency of snow removal per month; (C) number of time pork/food products enter the site per month; and (D) number of manure or effluent removals per month. Higher partial dependence values, represented by the y-axis, indicate a large marginal effect of practice or demographic. The x-axis represents the observed values of the practice or demographic.

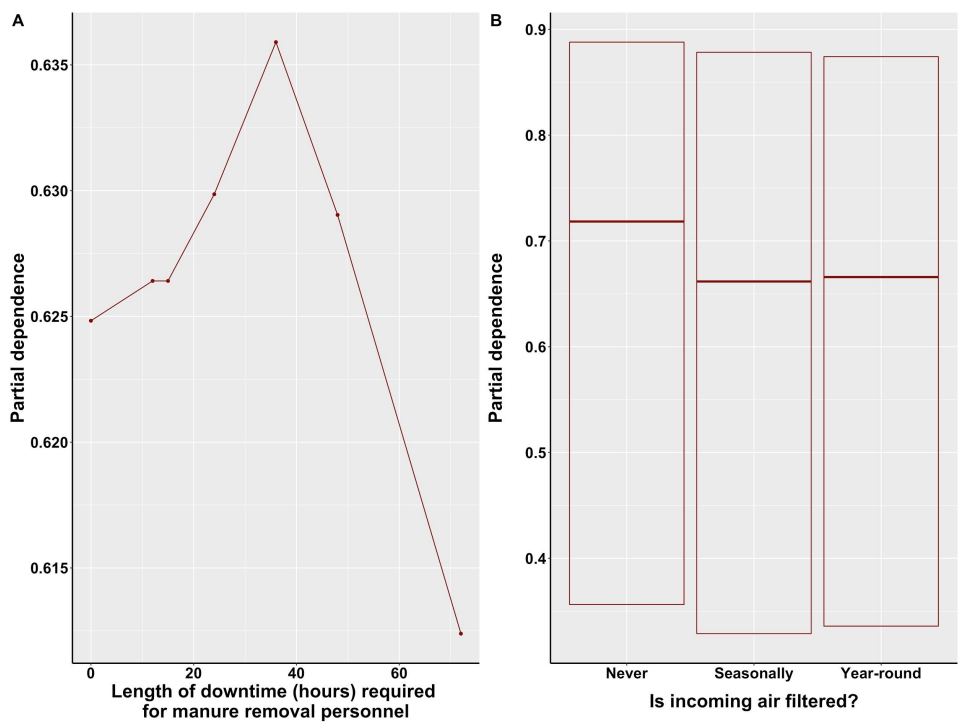

**Figure S11. Partial dependence plots showing the marginal effect of biosecurity practices and farm demographics on the model.** The following biosecurity practices and farm demographics are presented here: (A) length of downtime required for manure removal personnel; and (B) whether incoming air is filtered. Higher partial dependence values, represented by the y-axis, indicate a large marginal effect of practice or demographic. The x-axis represents the observed values of the practice or demographic.

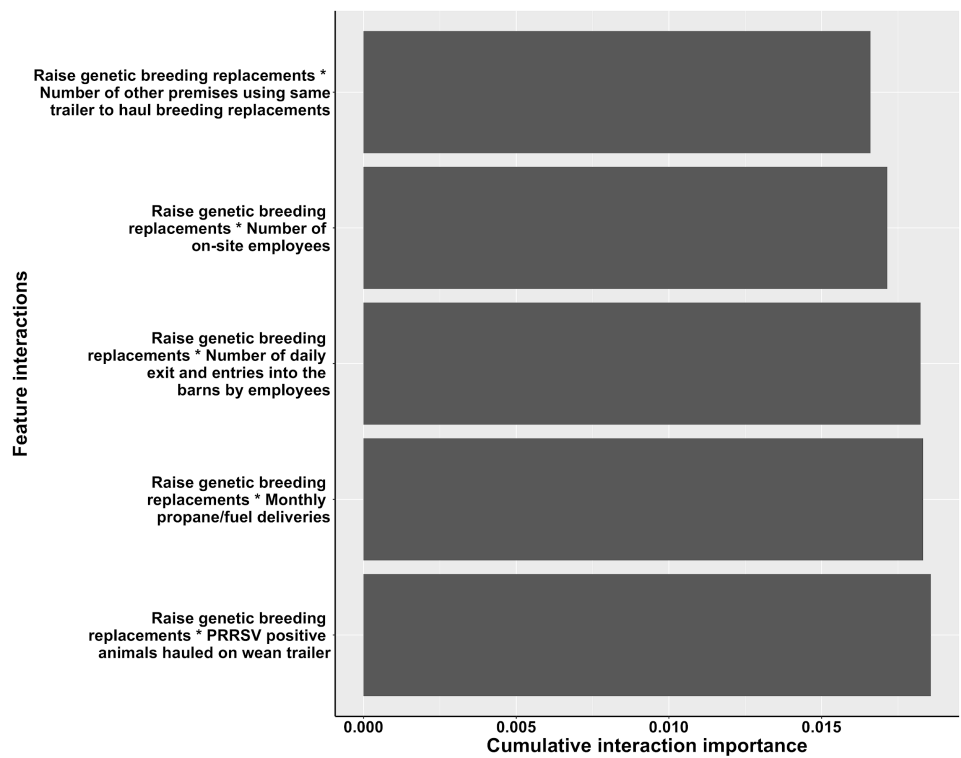

**Figure S12. Cumulative interaction importance of biosecurity practices and farm demographics.** Interaction strengths were calculated through Friedmans H statistic and ranked according to importance by the function *MrIMLInteractions*. Raising genetic breeding replacements (i.e. being a multiplication site) was identified to interact with: PRRSV positive animals being hauled on the same trailer used for weaned animals; monthly propane/fuel deliveries; the number of daily entries and exits into the barns by employees; the number of on-site employees; and the number of other premises using the same trailer for hauling breeding replacements. Interactions are ordered by cumulative interaction importance (represented by the x-axis). Higher cumulative interaction importance values indicate a stronger importance of this interaction in the model.

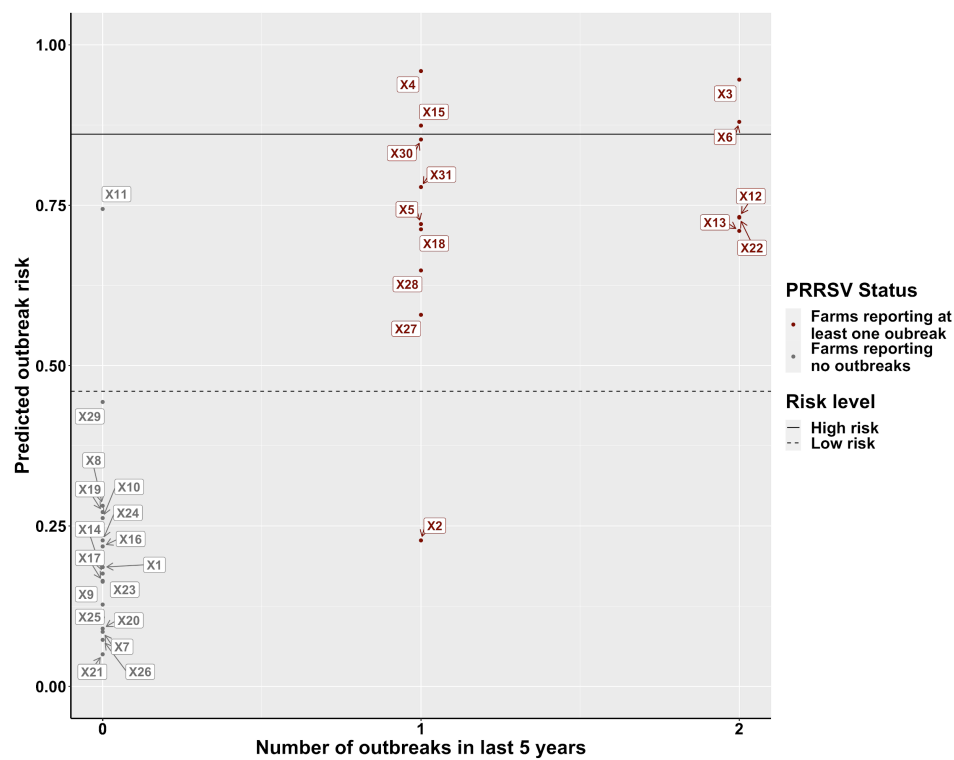

**Figure S13. Distribution of farms within a single production system example in relation to their predicted PRRSV outbreak risk and previous number of PRRSV outbreaks.** The x-axis represents the number of PRRSV outbreaks the farm has experienced in the previous five years -- red indicates at least one previous outbreak and grey indicates no previous outbreaks. The y-axis represents the predicted PRRSV outbreak risk produced by the machine learning model. Farms below the dashed line are considered low risk, between the dashed line and the solid line are considered medium risk, and above the solid line are considered high risk.

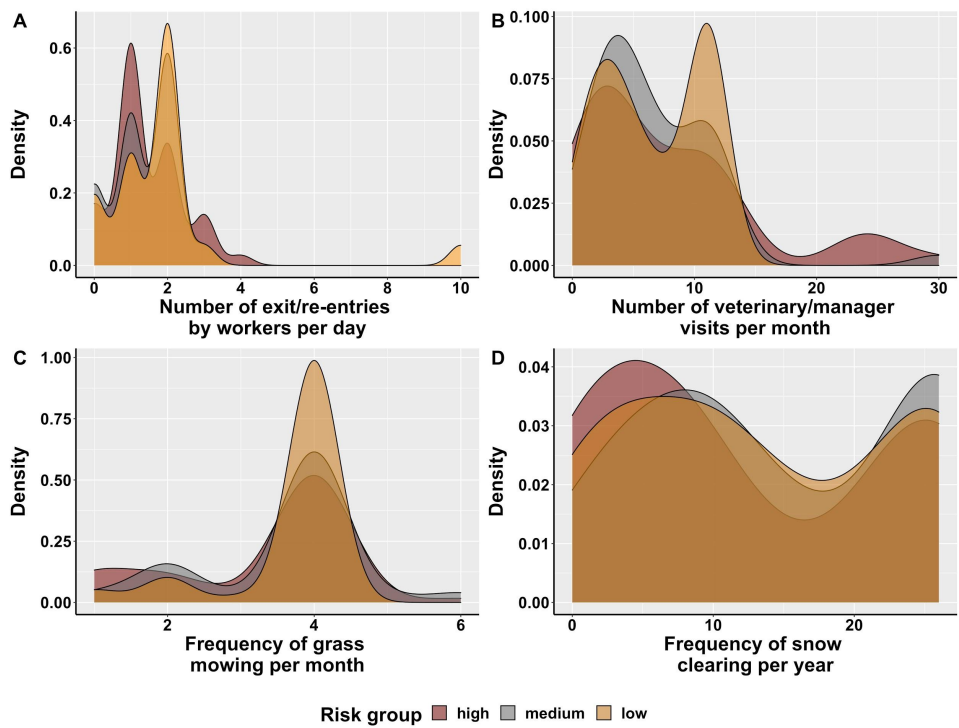

**Figure S14. Distribution of observed biosecurity practice and farm demographic values by PRRSV outbreak risk category.** The predicted PRRSV outbreak risk values were categorized by a *tidymodels* discretization method to create categories for low, medium, and high predicted outbreak risk (Kuhn and Wickham, 2021). Distributions for most important biosecurity practices and farm demographics, identified by the Gini Index, are presented here: (A) number of exits and re-entries by workers per day (p = 0.18); (B) the number of veterinarian/manager visits per month (p = 0.48); (C) frequency of grass mowing per month (p = 0.24); and (D) frequency of snow clearing per year (p = 0.42).

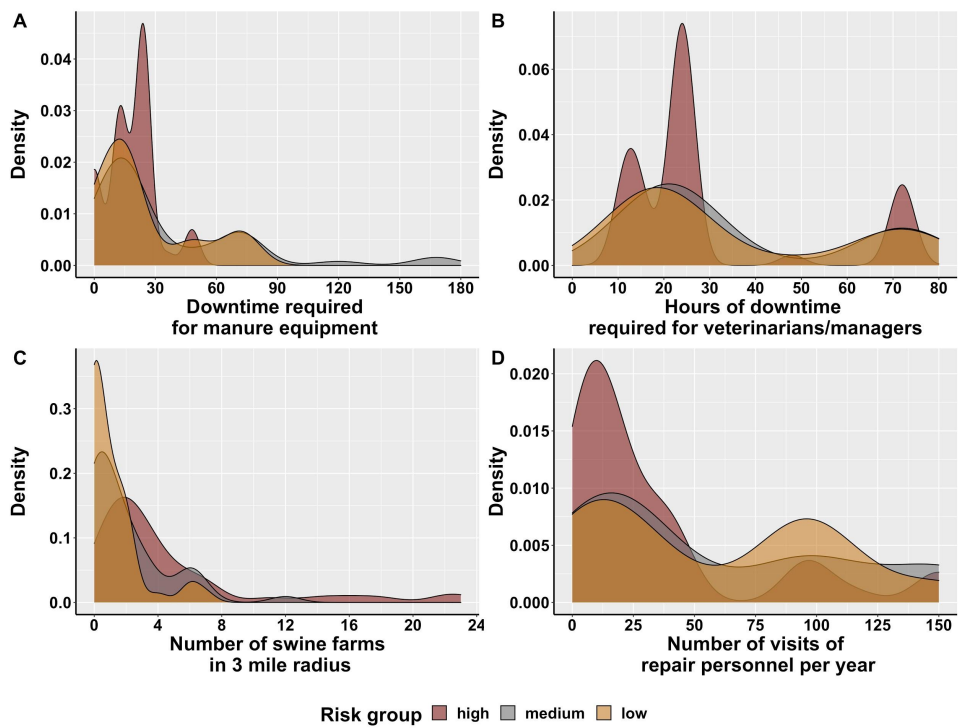

**Figure S15. Distribution of observed biosecurity practice and farm demographic values by PRRSV outbreak risk category.** The predicted PRRSV outbreak risk values were categorized by a *tidymodels* discretization method to create categories for low, medium, and high predicted outbreak risk (Kuhn and Wickham, 2021). Distributions for most important biosecurity practices and farm demographics, identified by the Gini Index, are presented here: (A) downtime required for manure equipment (p < 0.05); (B) downtime required for veterinarians and managers (p = 0.41); (C) number of swine farms in a three-mile radius (p < 0.000); and (D) number of repair personnel visits per year (p < 0.05).

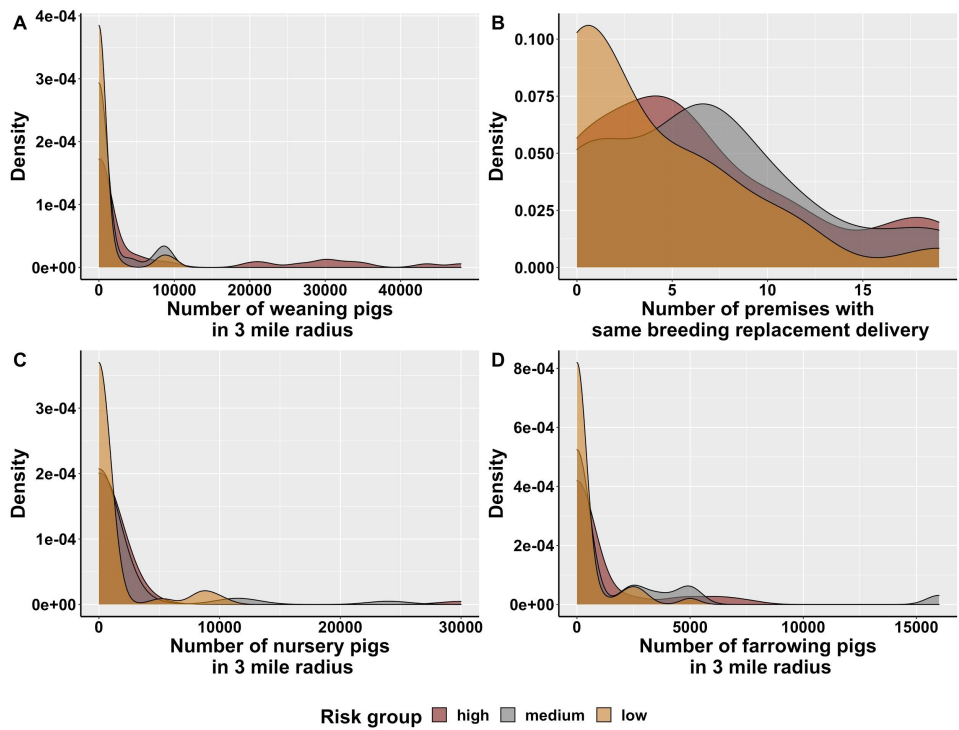

**Figure S16. Distribution of observed biosecurity practice and farm demographic values by PRRSV outbreak risk category.** The predicted PRRSV outbreak risk values were categorized by a *tidymodels* discretization method to create categories for low, medium, and high predicted outbreak risk (Kuhn and Wickham, 2021). Distributions for most important biosecurity practices and farm demographics, identified by the Gini Index, are presented here: (A) number of weaned pigs in a three-mile radius ($p < 0.000$); (B) number of premises with the same breeding replacement delivery ($p < 0.05$); (C) number of nursery pigs in a three-mile radius ($p = 0.79$); and (D) number of farrowing pigs in a three-mile radius ($p = 0.16$).

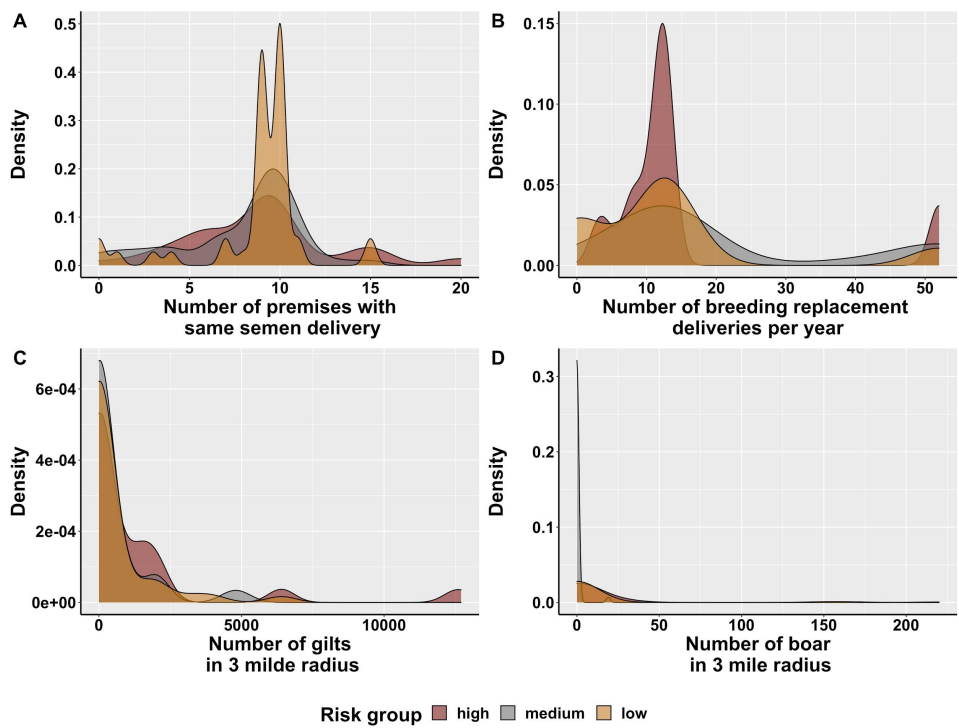

**Figure S17. Distribution of observed biosecurity practice and farm demographic values by PRRSV outbreak risk category.** The predicted PRRSV outbreak risk values were categorized by a *tidymodels* discretization method to create categories for low, medium, and high predicted outbreak risk (Kuhn and Wickham, 2021). Distributions for most important biosecurity practices and farm demographics, identified by the Gini Index, are presented here: (A) number of premises with same semen delivery (p = 0.58); (B) the number of breeding replacement deliveries per year (p = 0.06); (C) number of gilts in a three-mile radius (p = 0.06); and (D) number of boar in a three mile radius (p = 0.36).

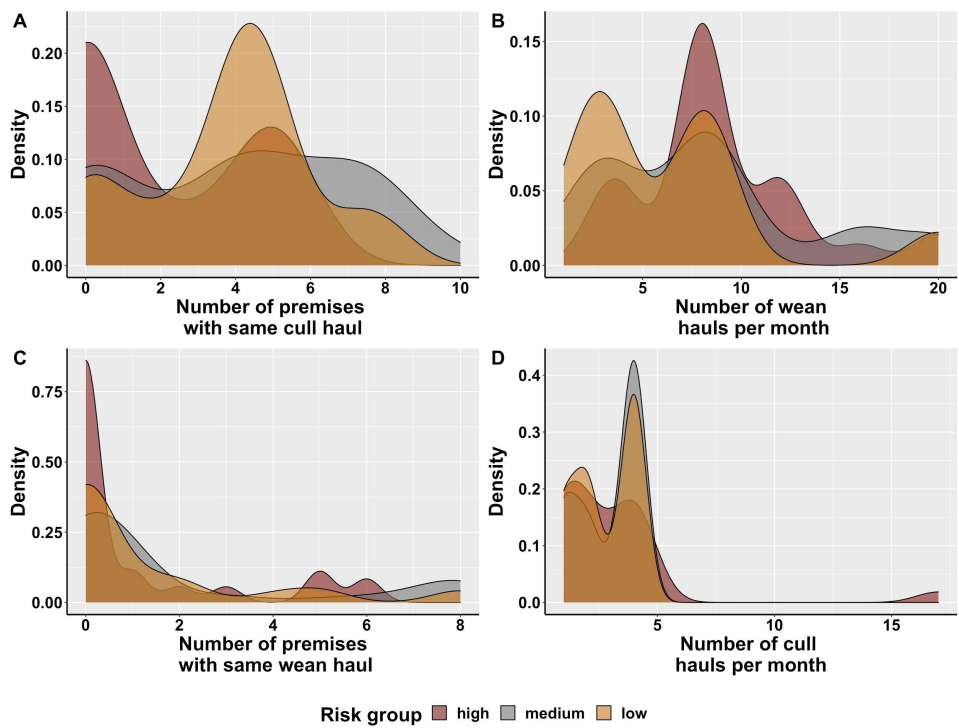

**Figure S18. Distribution of observed biosecurity practice and farm demographic values by PRRSV outbreak risk category.** The predicted PRRSV outbreak risk values were categorized by a *tidymodels* discretization method to create categories for low, medium, and high predicted outbreak risk (Kuhn and Wickham, 2021). Distributions for most important biosecurity practices and farm demographics, identified by the Gini Index, are presented here: (A) number of premises with the same trailer for hauling cull animals (p < 0.01); (B) number of wean hauls per month (p < 0.05); (C) number of premises with the same trailer to haul weaned animals (p = 0.26); and (D) number of cull hauls per month (p = 0.74).

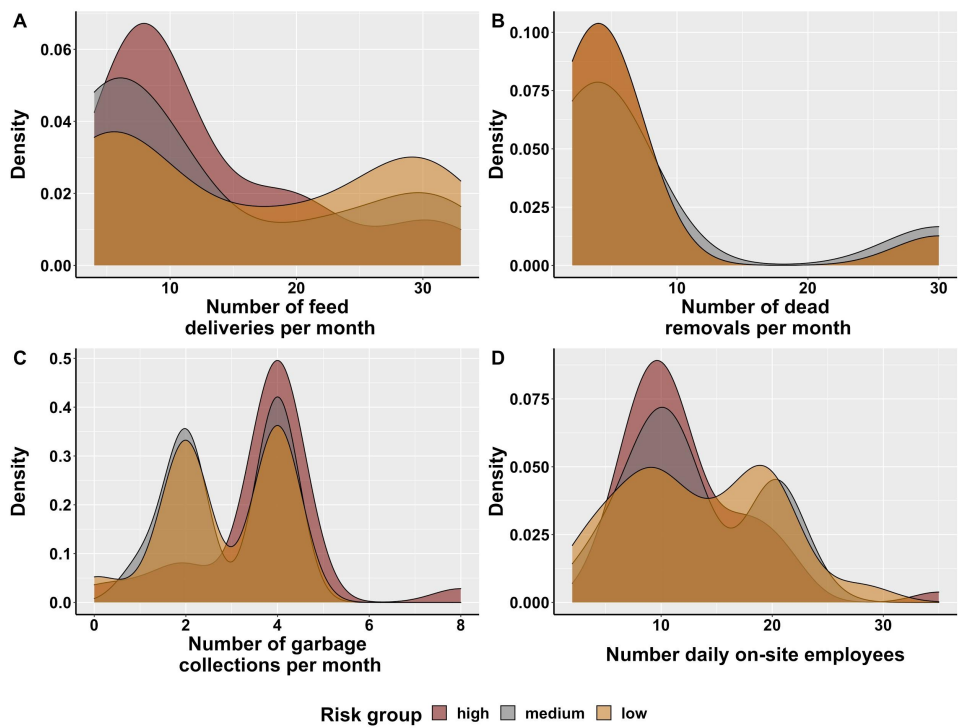

**Figure S19. Distribution of observed biosecurity practice and farm demographic values by PRRSV outbreak risk category.** The predicted PRRSV outbreak risk values were categorized by a *tidymodels* discretization method to create categories for low, medium, and high predicted outbreak risk (Kuhn and Wickham, 2021). Distributions for most important biosecurity practices and farm demographics, identified by the Gini Index, are presented here: (A) number of feed deliveries per month (p = 0.18); (B) the number of dead removals per month (p = 0.57); (C) number of garbage collections per month (p < 0.01); and (D) number of employees on-site daily (p = 0.7).

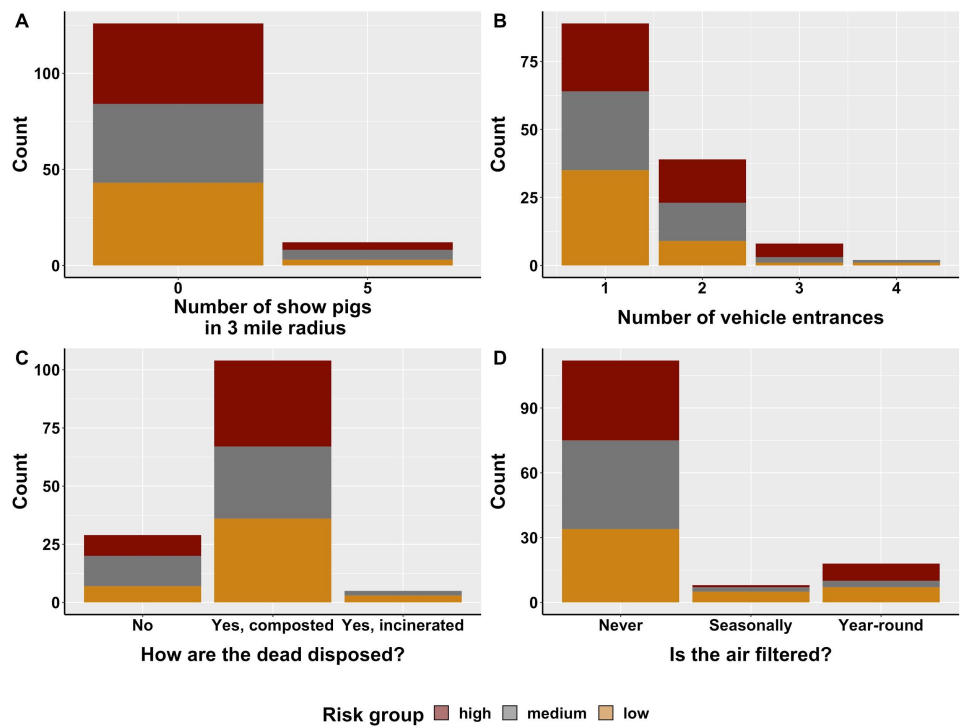

**Figure S20. Distribution of observed biosecurity practice and farm demographic values by PRRSV outbreak risk category.** The predicted PRRSV outbreak risk values were categorized by a *tidymodels* discretization method to create categories for low, medium, and high predicted outbreak risk (Kuhn and Wickham, 2021). Distributions for most important biosecurity practices and farm demographics, identified by the Gini Index, are presented here: (A) number of show pigs in a three-mile radius (p = 0.99); (B) number of vehicle entrances (p = 0.2); (C) whether dead are disposed in site and which method is used (p = 0.25); and (D) whether incoming air is filtered (p = 0.29).

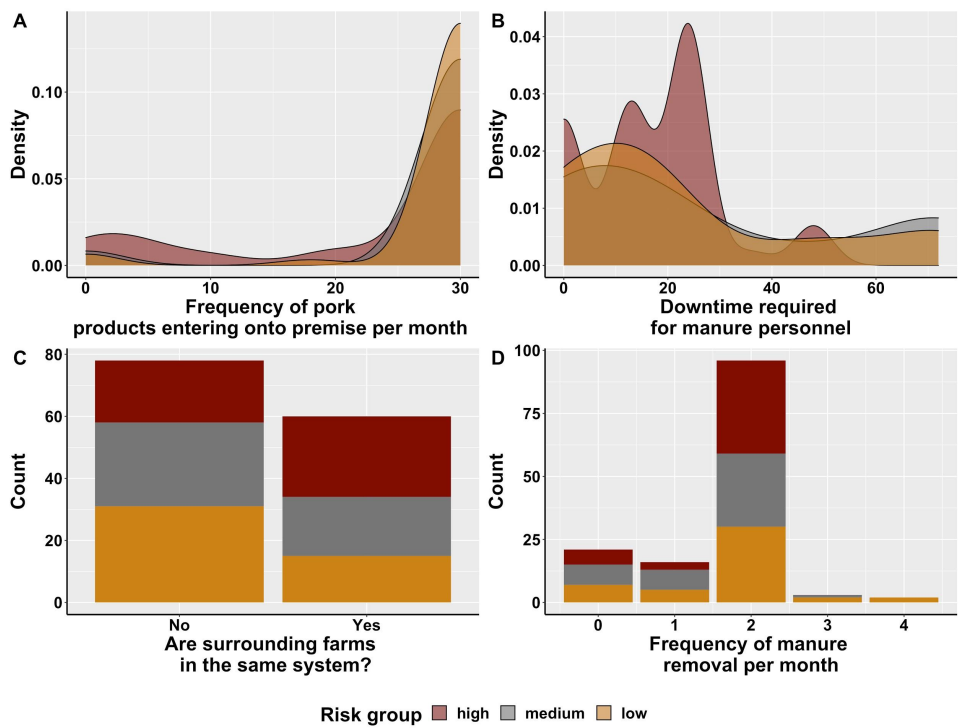

**Figure S21. Distribution of observed biosecurity practice and farm demographic values by PRRSV outbreak risk category.** The predicted PRRSV outbreak risk values were categorized by a *tidymodels* discretization method to create categories for low, medium, and high predicted outbreak risk (Kuhn and Wickham, 2021). Distributions for most important biosecurity practices and farm demographics, identified by the Gini Index, are presented here: (A) frequency of pork products entering the premises per month ($p < 0.01$); (B) downtime required for manure personnel ($p = 0.15$); (C) whether the surrounding farms belong to the same system ($p = 0.05$); and (D) frequency of manure removals per month ($p = 0.66$).

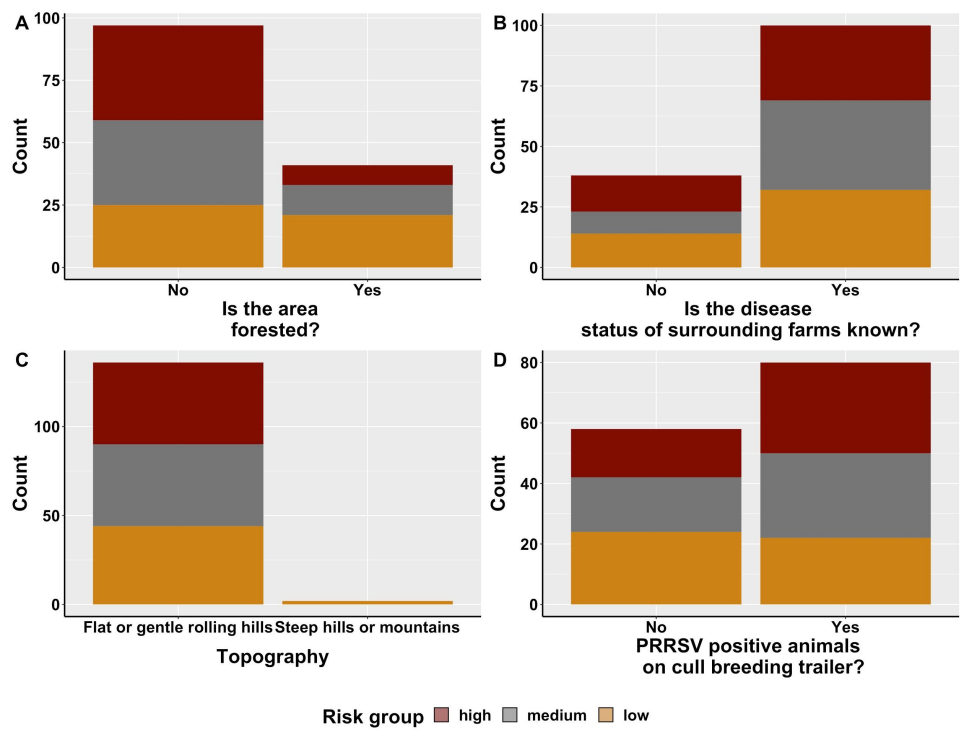

**Figure S22. Distribution of observed biosecurity practice and farm demographic values by PRRSV outbreak risk category.** The predicted PRRSV outbreak risk values were categorized by a *tidymodels* discretization method to create categories for low, medium, and high predicted outbreak risk (Kuhn and Wickham, 2021). Distributions for most important biosecurity practices and farm demographics, identified by the Gini Index, are presented here: (A) whether the surrounding area is forested (p < 0.05); (B) whether the disease status of the surrounding farms is known (p = 0.29); (C) the topography of the surrounding area (p = 0.14); and (D) whether PRRSV positive animals are hauled on the cull trailer (p = 0.14).

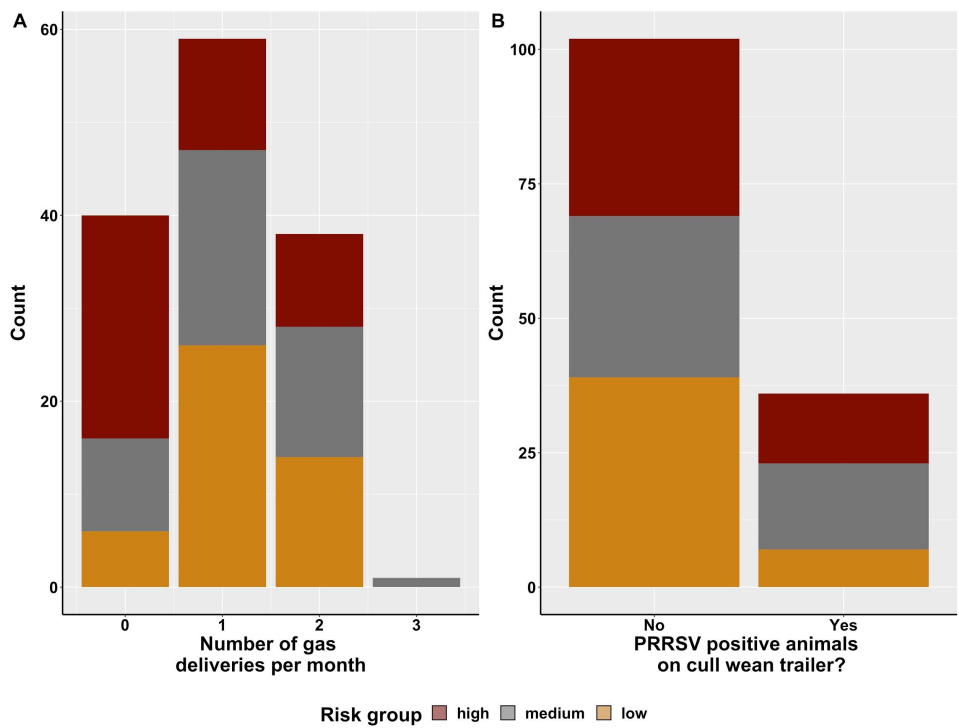

**Figure S23. Distribution of observed biosecurity practice and farm demographic values by PRRSV outbreak risk category.** The predicted PRRSV outbreak risk values were categorized by a *tidymodels* discretization method to create categories for low, medium, and high predicted outbreak risk (Kuhn and Wickham, 2021). Distributions for most important biosecurity practices and farm demographics, identified by the Gini Index, are presented here: (A) the number of gas/fuel deliveries per month (p < 0.01); and (B) whether PRRSV positive animals are hauled on the cull wean trailer (p = 0.08).

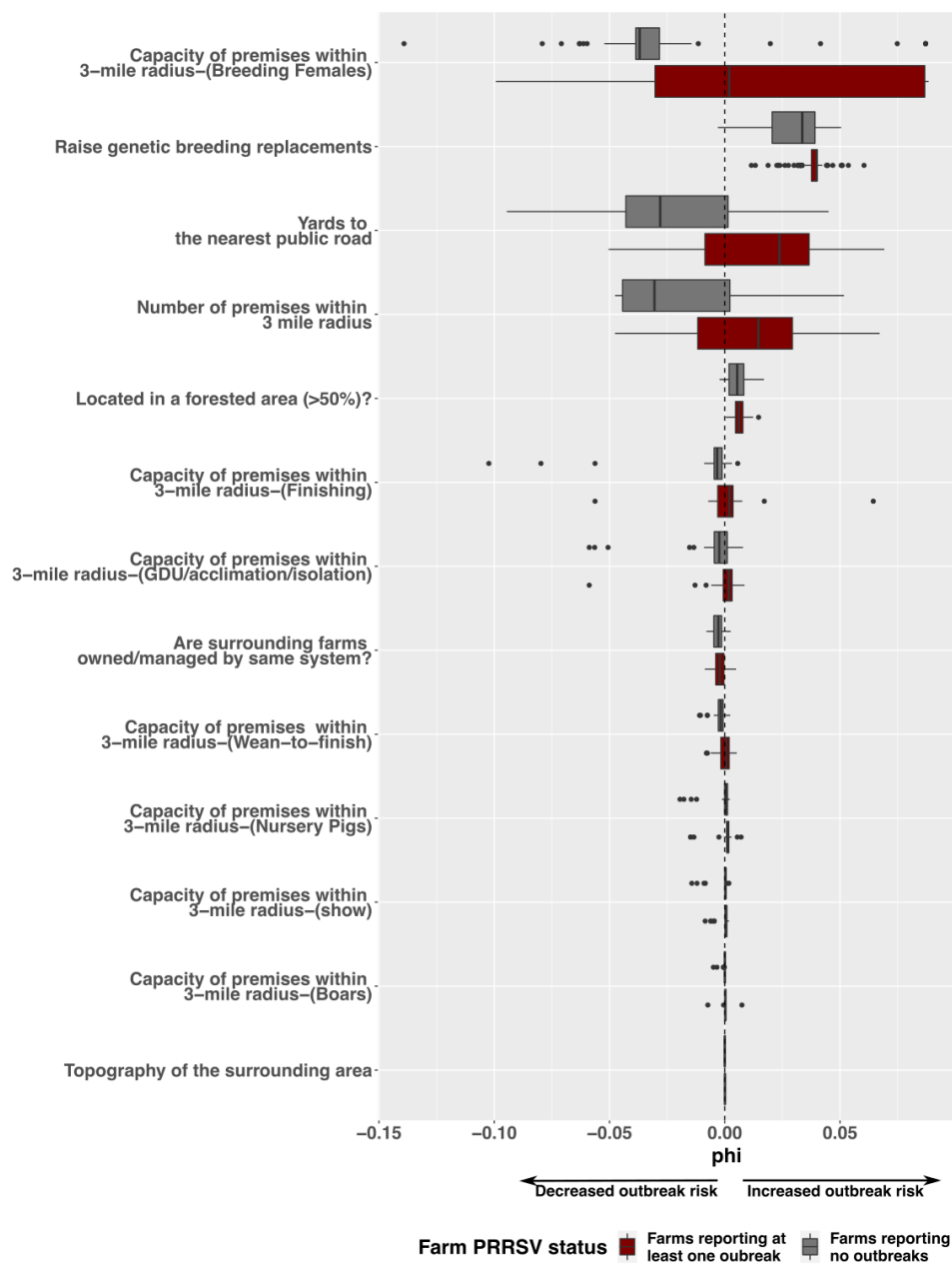

**Figure S24. The rank of farm demographics produced by local model agnostic interpretation methods.** The boxplots represent a summary of the breakDown values generated for each farm demographic per farm. The current status of farms is presented as having reported at least one outbreak (red) or having reported no outbreaks (gray). The axis represents the contribution of the demographic to PRRSV outbreak prediction. Values above zero (right of the dashed line) are contributing to an increased risk prediction and values below zero (left of the dashed line) are contributing to a decreased risk prediction. Demographics are ordered in relation to their absolute mean contribution.

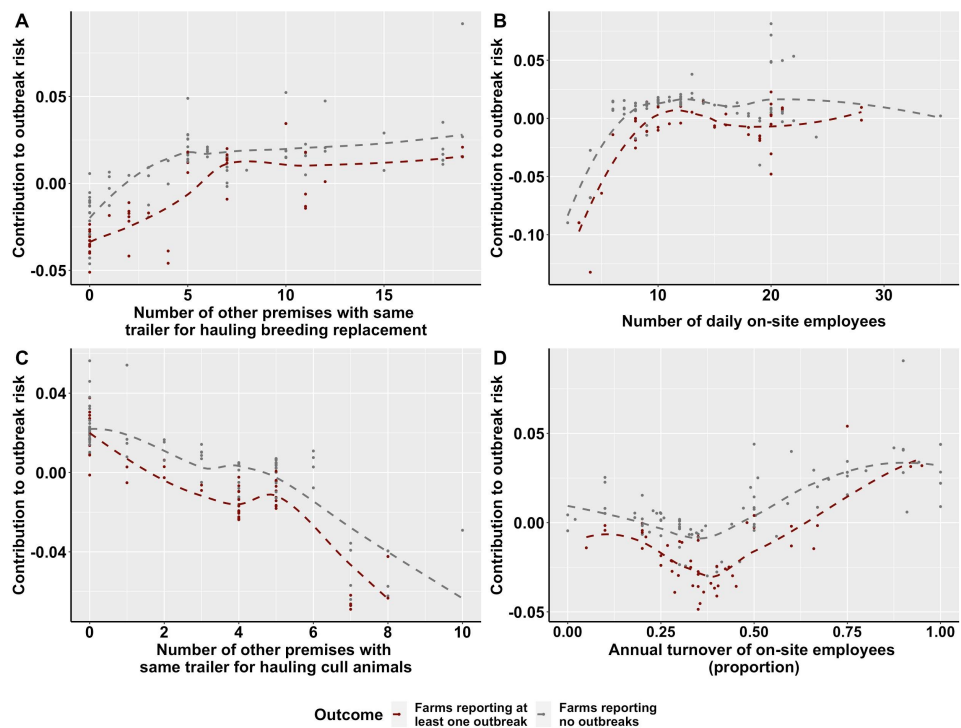

**Figure S25. Dispersal plots showing the directionality of the contribution of the top four important biosecurity practices and farm demographics on predicted PRRSV outbreak risk at individual farms.** The following biosecurity practices and farm demographics are presented here: (A) Number of other premises using the same trailer for hauling breeding replacements; (B) number of daily on-site employees; (C) number of other premises using the same trailer to haul cull animals; and (D) annual turnover of on-site employees. Contribution values, represented by the y-axis, indicate the contribution of the practice or demographic on predicted PRRSV risk. Values above zero are contributing to an increased risk prediction and values below zero are contributing to a decreased risk prediction. The x-axis represents the observed values of the practice or demographic.

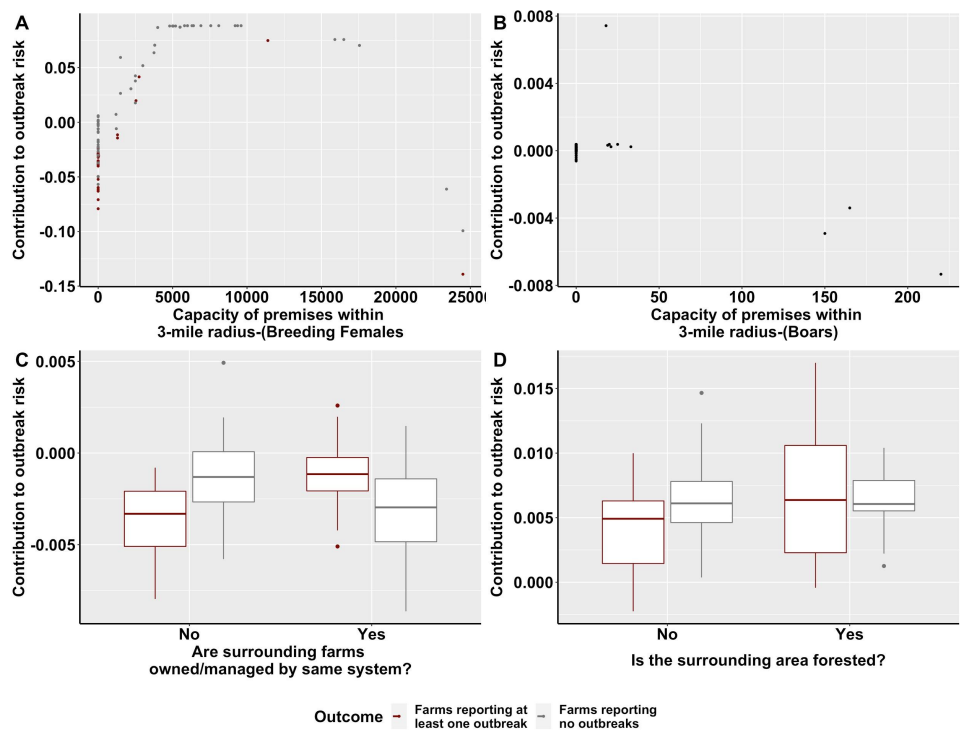

**Figure S26. Dispersal plots showing the directionality of the contribution of biosecurity practices and farm demographics on predicted PRRSV outbreak risk at individual farms.** The following biosecurity practices and farm demographics are presented here: (A) capacity of breeding females within a three mile radius; (B) capacity of boars within a three mile radius; (C) whether surrounding farms owned by the same production system; and (D) whether the surrounding area is forested. Contribution values, represented by the y-axis, indicate the contribution of the practice or demographic on predicted PRRSV risk. Values above zero are contributing to an increased risk prediction and values below zero are contributing to a decreased risk prediction. The x-axis represents the observed values of the practice or demographic.

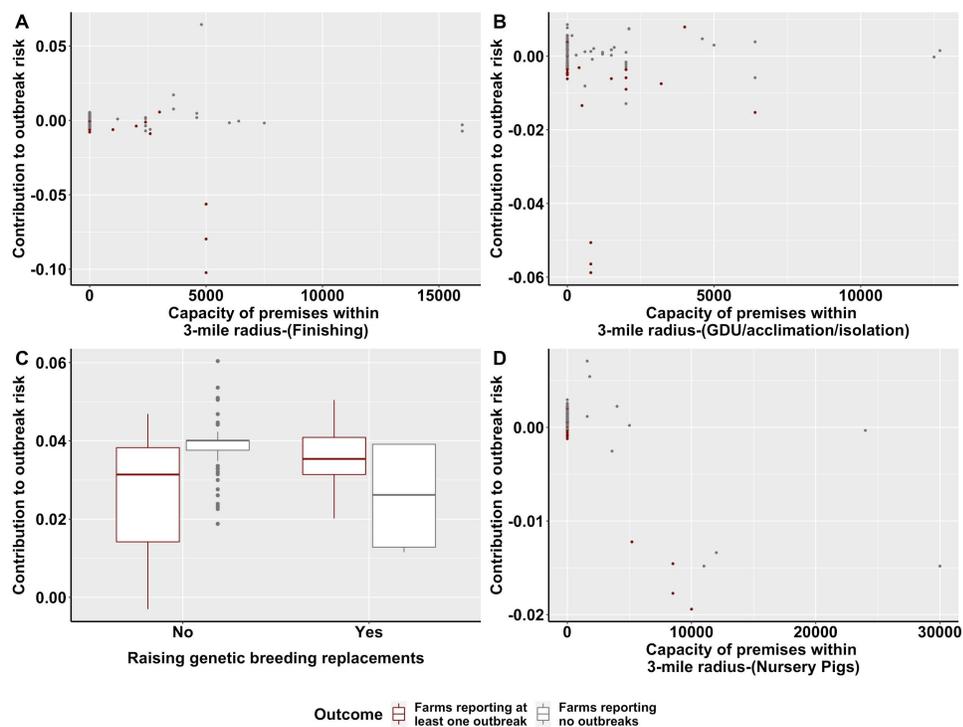

**Figure S27. Dispersal plots showing the directionality of the contribution of biosecurity practices and farm demographics on predicted PRRSV outbreak risk at individual farms.** The following biosecurity practices and farm demographics are presented here: (A) capacity of finishing pigs within a three mile radius; (B) capacity of gilts within a three mile radius; (C) whether farm is primarily raising genetic breeding replacements; and (D) capacity of nursery pigs within a three mile radius. Contribution values, represented by the y-axis, indicate the contribution of the practice or demographic on predicted PRRSV risk. Values above zero are contributing to an increased risk prediction and values below zero are contributing to a decreased risk prediction. The x-axis represents the observed values of the practice or demographic.

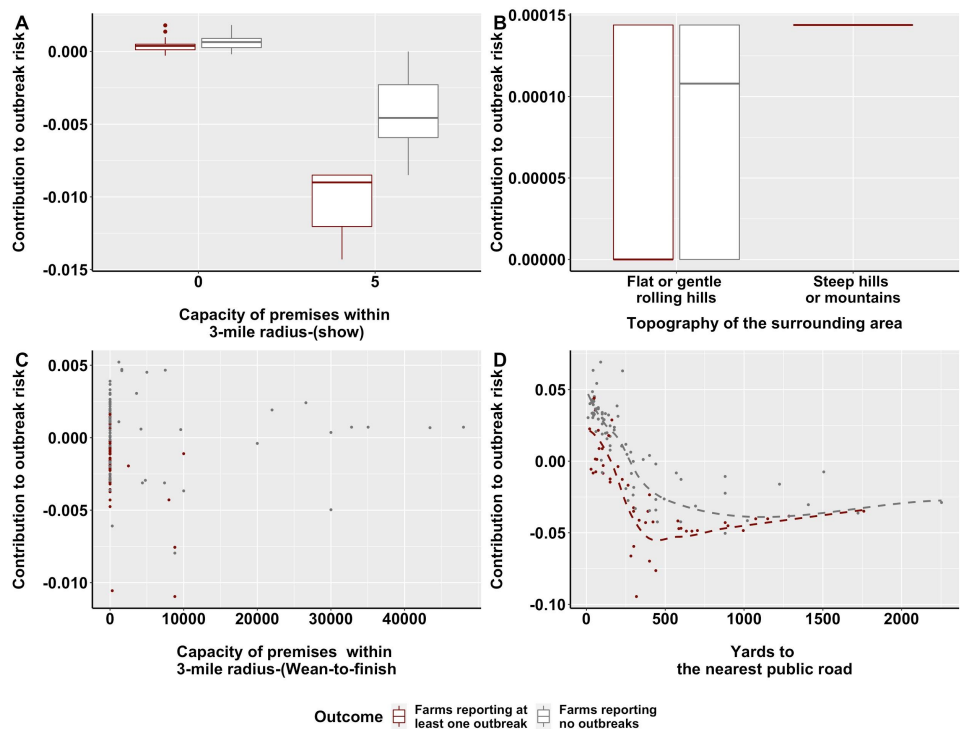

**Figure S28. Dispersal plots showing the directionality of the contribution of biosecurity practices and farm demographics on predicted PRRSV outbreak risk at individual farms.** The following biosecurity practices and farm demographics are presented here: (A) capacity of show pigs within a three mile radius; (B) topography of the surrounding area; (C) capacity of wean-to-finishers within a three mile radius; and (D) yards to nearest public road. Contribution values, represented by the y-axis, indicate the contribution of the practice or demographic on predicted PRRSV risk. Values above zero are contributing to an increased risk prediction and values below zero are contributing to a decreased risk prediction. The x-axis represents the observed values of the practice or demographic.

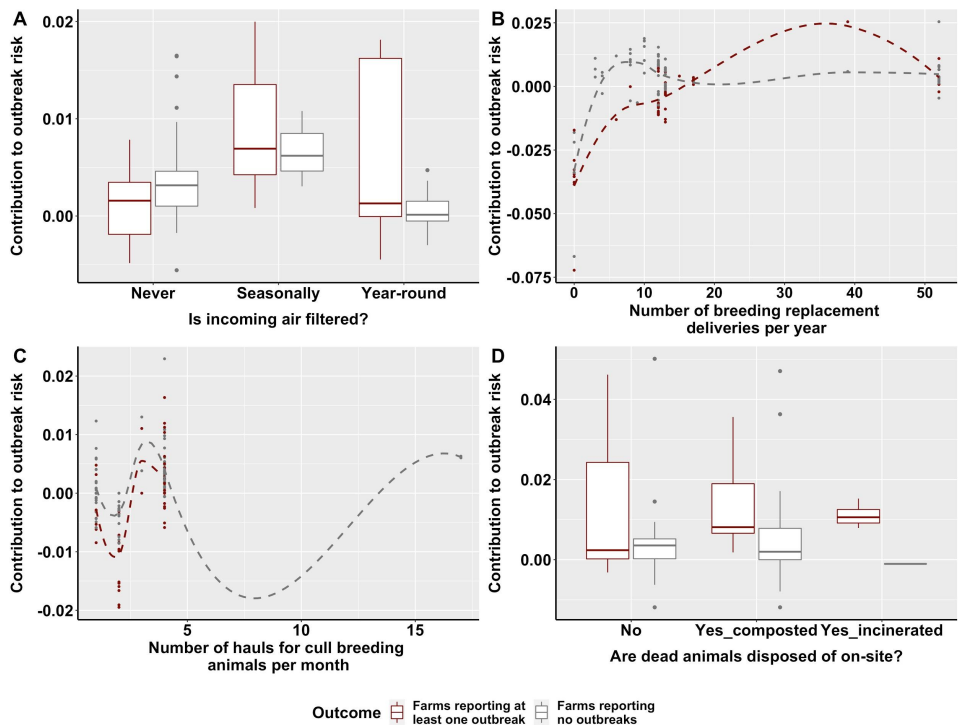

**Figure S29. Dispersal plots showing the directionality of the contribution of biosecurity practices and farm demographics on predicted PRRSV outbreak risk at individual farms.** The following biosecurity practices and farm demographics are presented here: (A) whether the incoming air is filtered; (B) number of breeding replacements deliveries per year; (C) number of hauls for cull breeding animals per month; and (D) whether animals are disposed on-site and the method used. Contribution values, represented by the y-axis, indicate the contribution of the practice or demographic on predicted PRRSV risk. Values above zero are contributing to an increased risk prediction and values below zero are contributing to a decreased risk prediction. The x-axis represents the observed values of the practice or demographic.

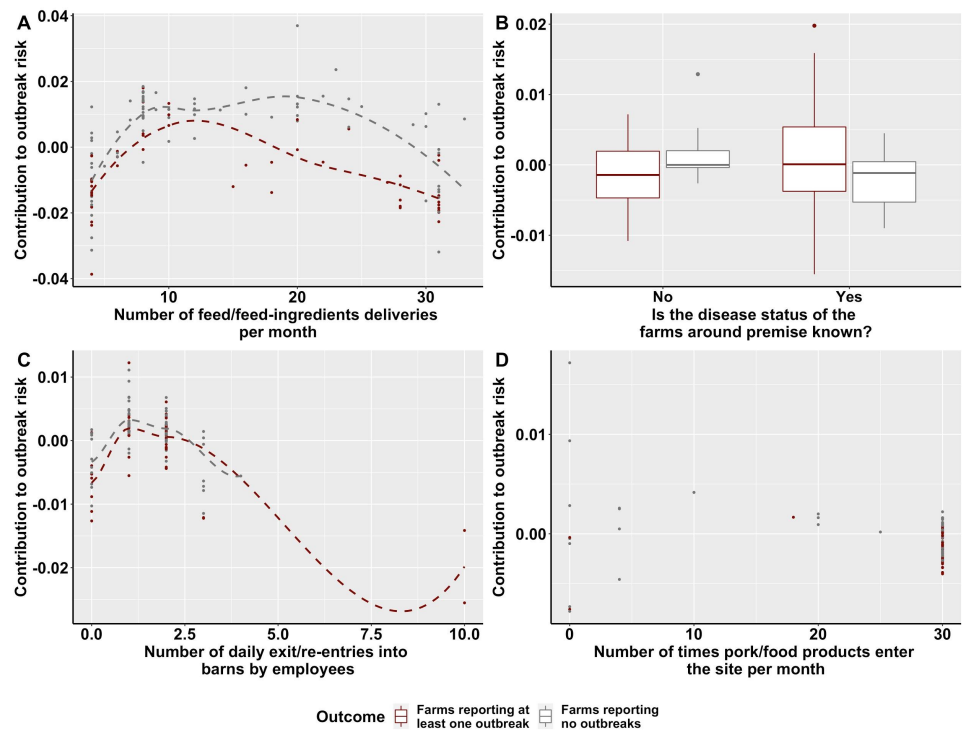

**Figure S30. Dispersal plots showing the directionality of the contribution of biosecurity practices and farm demographics on predicted PRRSV outbreak risk at individual farms.** The following biosecurity practices and farm demographics are presented here: (A) number of feed/feed-ingredient deliveries per month; (B) whether the disease status of the surrounding farms is known; (C) number of daly exit and reentries into the barns by employees; and (D) number of times pork/food products enter the site per month. Contribution values, represented by the y-axis, indicate the contribution of the practice or demographic on predicted PRRSV risk. Values above zero are contributing to an increased risk prediction and values below zero are contributing to a decreased risk prediction. The x-axis represents the observed values of the practice or demographic.

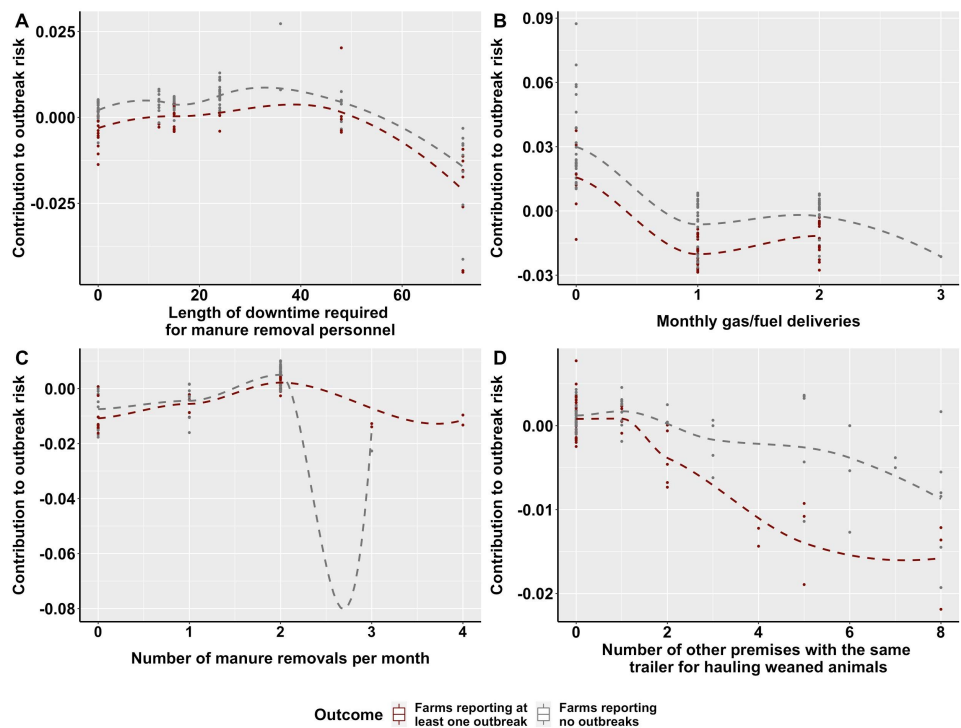

**Figure S31. Dispersal plots showing the directionality of the contribution of biosecurity practices and farm demographics on predicted PRRSV outbreak risk at individual farms.** The following biosecurity practices and farm demographics are presented here: (A) length of downtime required for manure removal personnel; (B) monthly gas/fuel deliveries; (C) number of manure removals per month; and (D) number of other premises with the same trailer for hauling weaned animals. Contribution values, represented by the y-axis, indicate the contribution of the practice or demographic on predicted PRRSV risk. Values above zero are contributing to an increased risk prediction and values below zero are contributing to a decreased risk prediction. The x-axis represents the observed values of the practice or demographic.

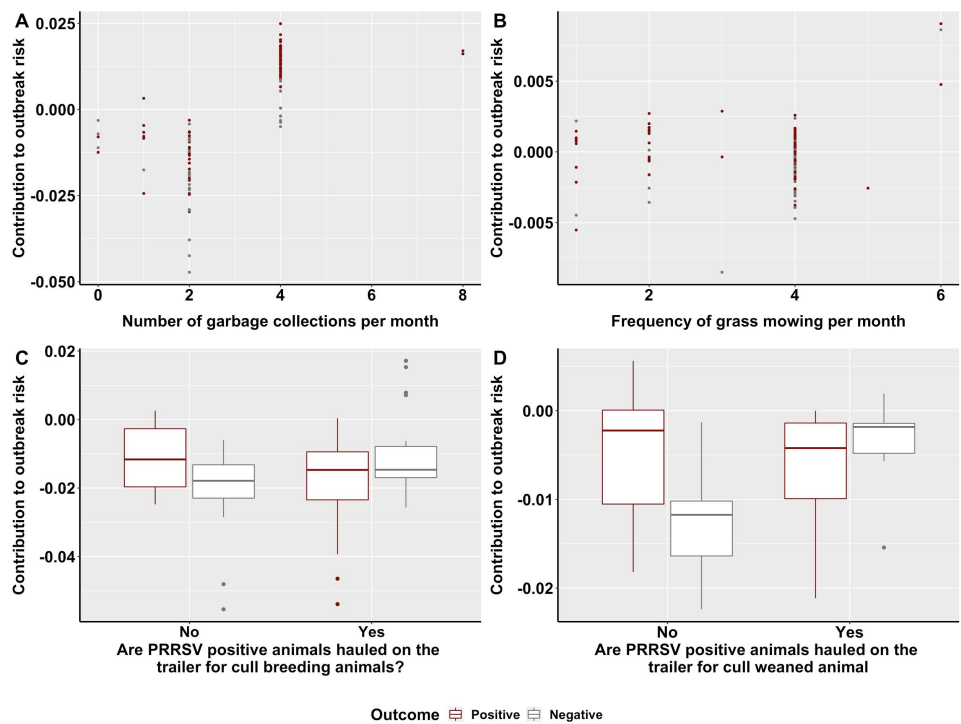

**Figure S32. Dispersal plots showing the directionality of the contribution of biosecurity practices and farm demographics on predicted PRRSV outbreak risk at individual farms.** The following biosecurity practices and farm demographics are presented here: (A) number of garbage collections per month; (B) frequency of grass mowing per month; (C) whether PRRSV positive animals are hauled on the cull breeding animal trailer; and (D) whether PRRSV positive animals are hauled on the cull wean animal trailer. Contribution values, represented by the y-axis, indicate the contribution of the practice or demographic on predicted PRRSV risk. Values above zero are contributing to an increased risk prediction and values below zero are contributing to a decreased risk prediction. The x-axis represents the observed values of the practice or demographic.

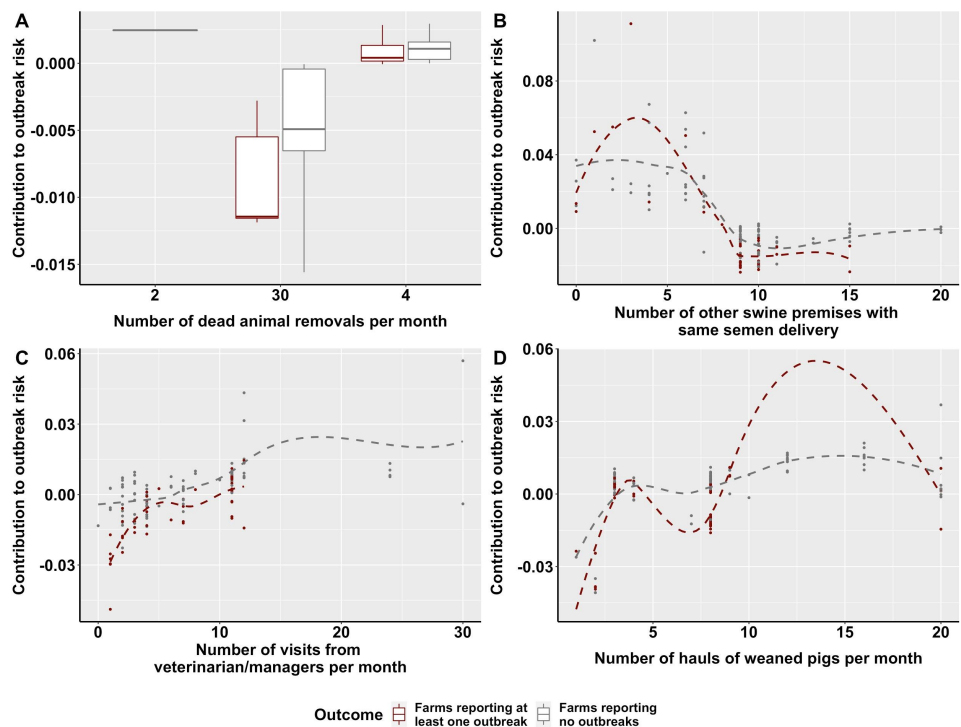

**Figure S33. Dispersal plots showing the directionality of the contribution of biosecurity practices and farm demographics on predicted PRRSV outbreak risk at individual farms.** The following biosecurity practices and farm demographics are presented here: (A) number of dead animal removals per month; (B) number of other swine premises with the same seme delivery; (C) number of visits from veterinarians and managers per month; and (D) number of hauls of weaned pigs per month. Contribution values, represented by the y-axis, indicate the contribution of the practice or demographic on predicted PRRSV risk. Values above zero are contributing to an increased risk prediction and values below zero are contributing to a decreased risk prediction. The x-axis represents the observed values of the practice or demographic.

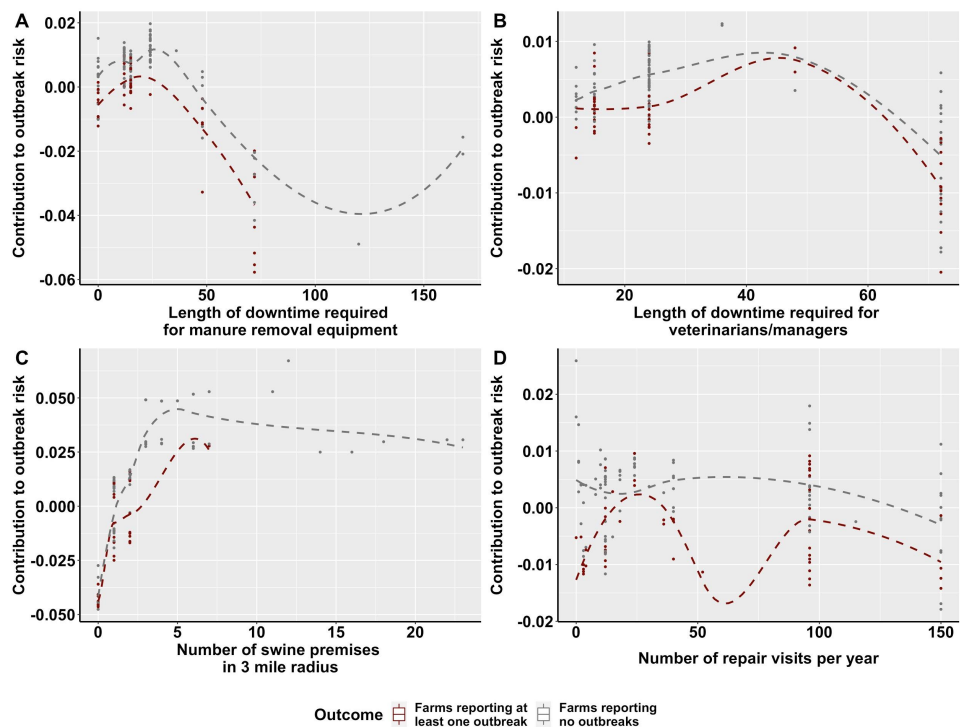

**Figure S34. Dispersal plots showing the directionality of the contribution of biosecurity practices and farm demographics on predicted PRRSV outbreak risk at individual farms.** The following biosecurity practices and farm demographics are presented here: (A) length of downtime required for manure removal equipment; (B) length of downtime required for veterinarians and managers; (C) number of swine premises in a three-mile radius; and (D) number of visits from repair personnel per year. Contribution values, represented by the y-axis, indicate the contribution of the practice or demographic on predicted PRRSV risk. Values above zero are contributing to an increased risk prediction and values below zero are contributing to a decreased risk prediction. The x-axis represents the observed values of the practice or demographic.

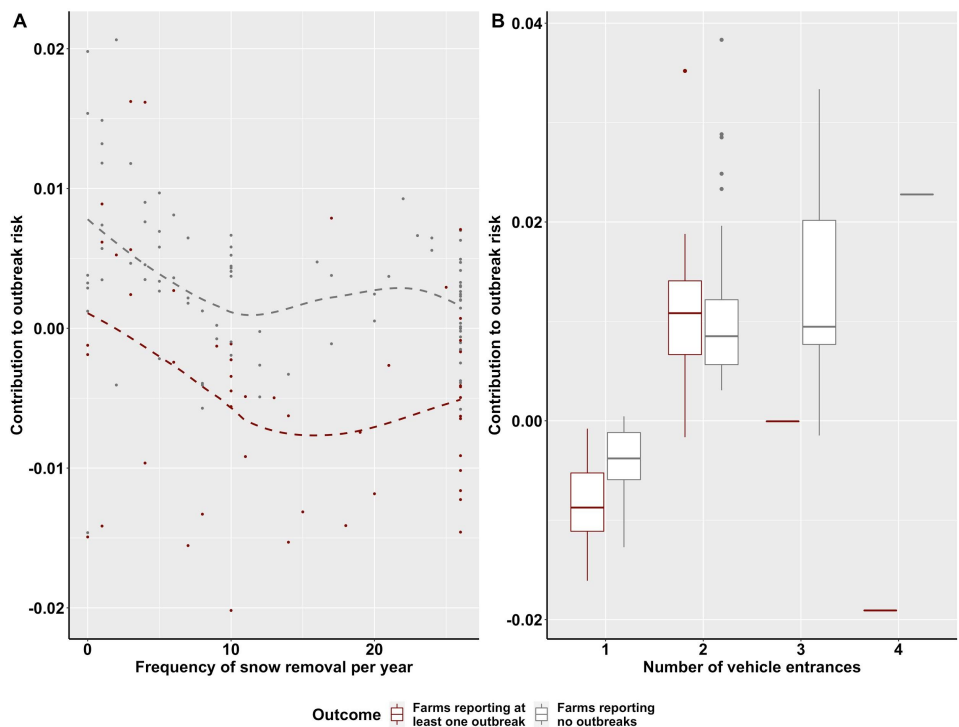

**Figure S35. Dispersal plots showing the directionality of the contribution of biosecurity practices and farm demographics on predicted PRRSV outbreak risk at individual farms.** The following biosecurity practices and farm demographics are presented here: (A) frequency of snow removal per year; and (B) number of vehicle entrances on the premises. Contribution values, represented by the y-axis, indicate the contribution of the practice or demographic on predicted PRRSV risk. Values above zero are contributing to an increased risk prediction and values below zero are contributing to a decreased risk prediction. The x-axis represents the observed values of the practice or demographic.

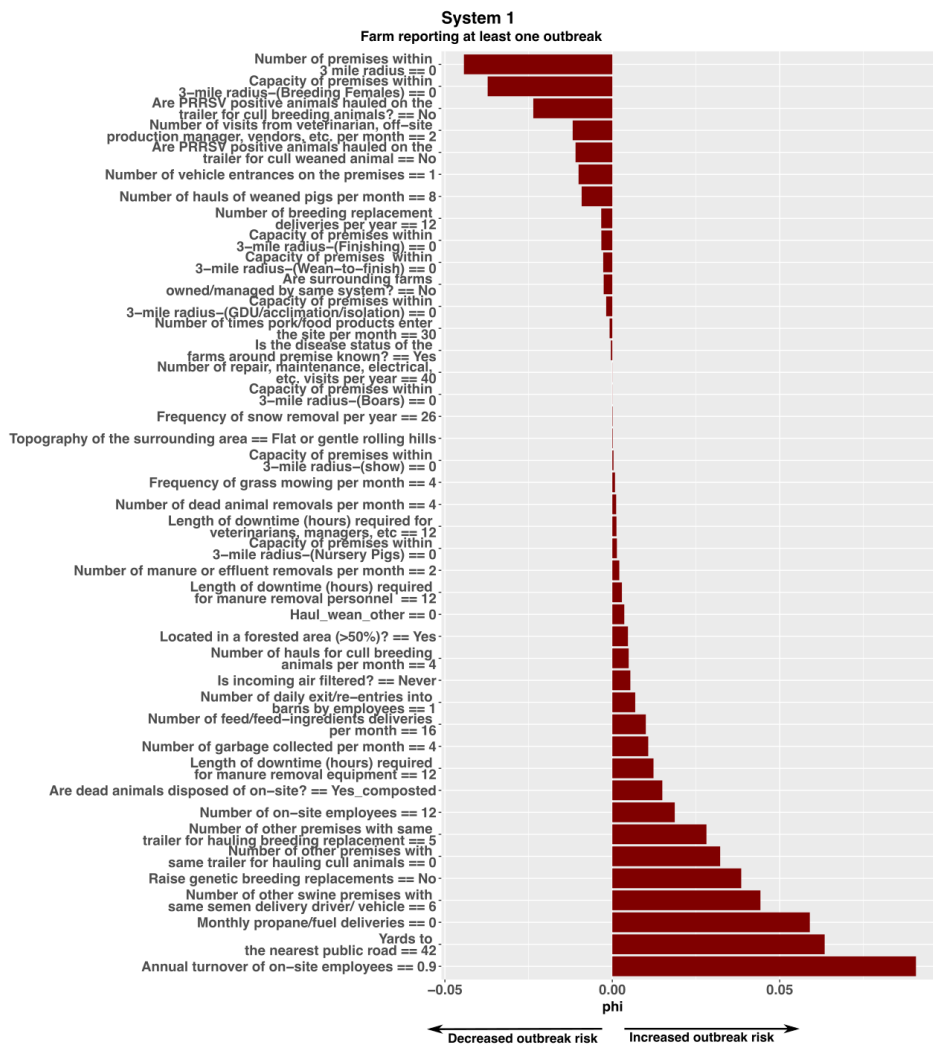

**Figure S36. The rank of biosecurity practices and farm demographics produced by local model agnostic interpretation methods for an individual farm reporting at least one outbreak.** Each bar on the waterfall plot represents the contribution of the biosecurity practice or farm demographic to the PRRSV predicted risk at the farm. The contribution value, phi, is represented by the axis. Values above zero (right of the dashed line) are contributing to an increased outbreak risk prediction and values below zero (left of the dashed line) are contributing to a decreased risk prediction.

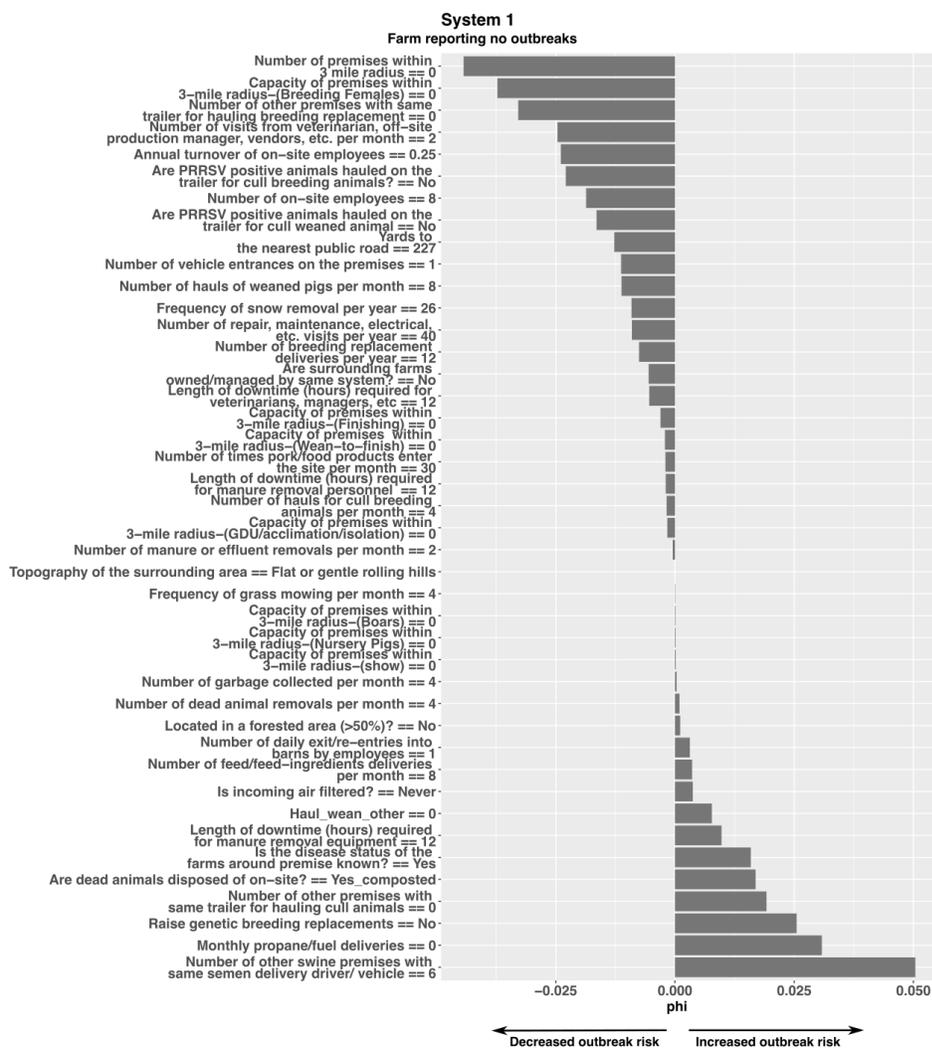

**Figure S37. The rank of biosecurity practices and farm demographics produced by local model agnostic interpretation methods for an individual farm reporting no outbreaks.** Each bar on the waterfall plot represents the contribution of the biosecurity practice or farm demographic to the PRRSV predicted risk at the farm. The contribution value, phi, is represented by the axis. Values above zero (right of the dashed line) are contributing to an increased outbreak risk prediction and values below zero (left of the dashed line) are contributing to a decreased risk prediction.

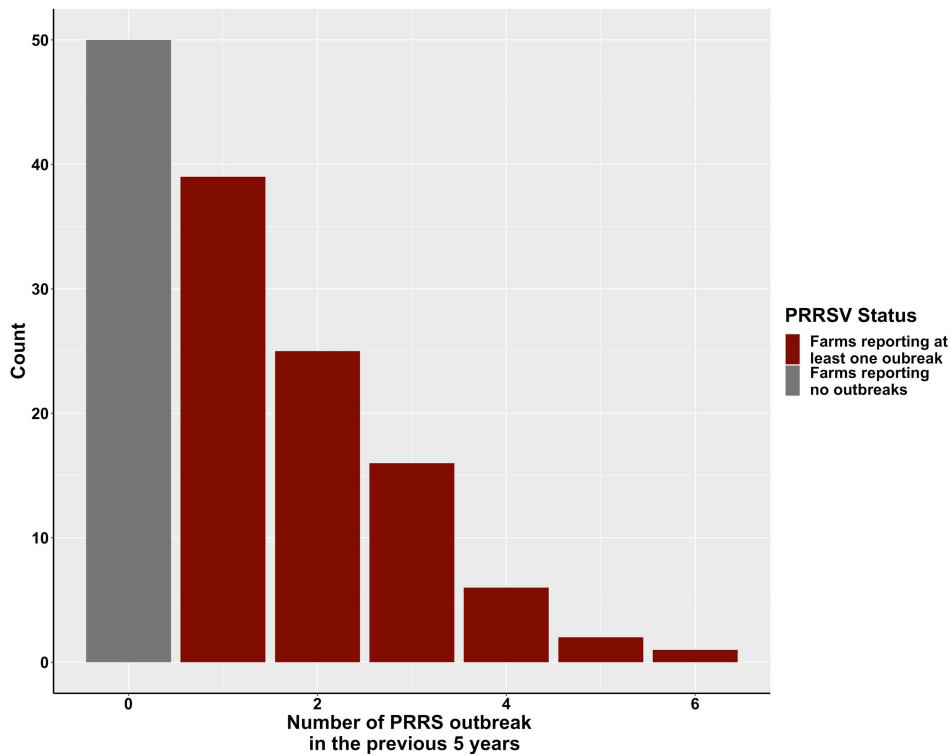

**Figure S38. Distribution of frequency of PRRSV outbreaks in the previous five years.** Each farm provided data regarding PRRSV outbreaks five years previous to the survey. Farms with no outbreaks in the previous five years and farms with at least one outbreak in the previous five years.

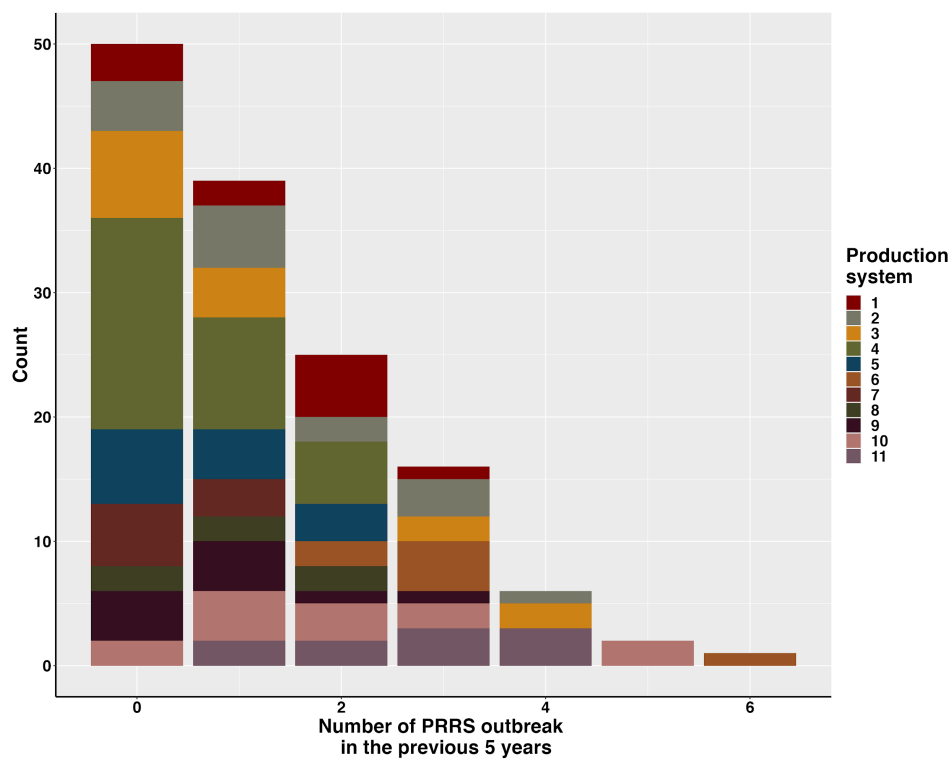

Figure 39. Distribution of frequency of PRRSV outbreaks in the previous five years by production system. Each farm provided data regarding PRRSV outbreaks five years previous to the survey.

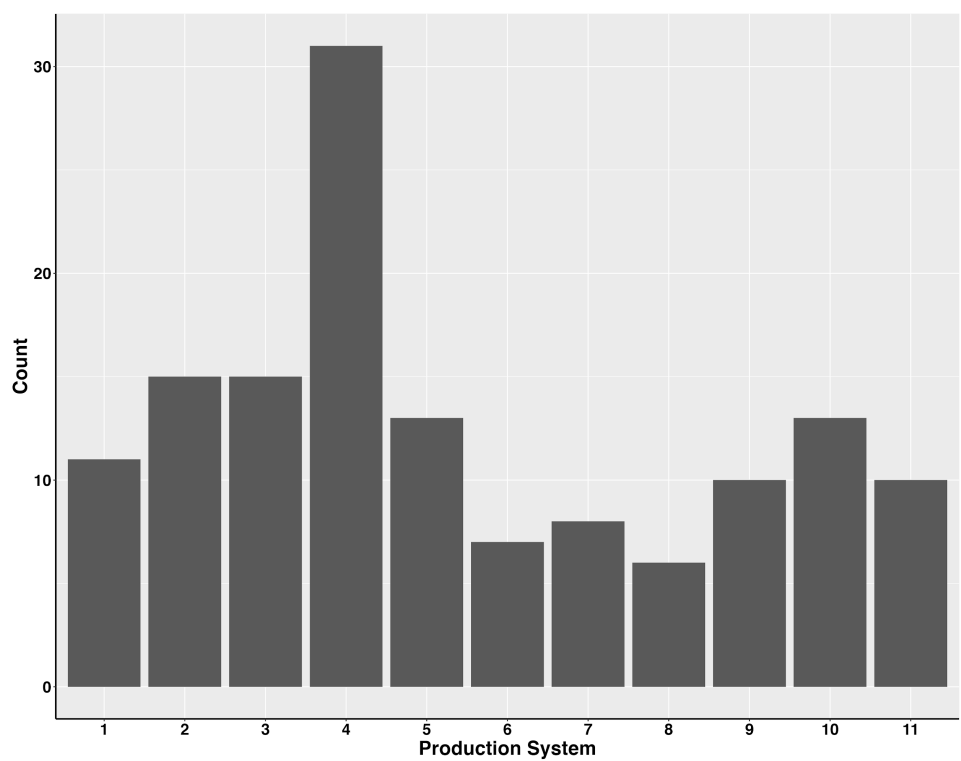

Figure S40. Distribution of farms among each production system. Participating production systems were asked to provide information for a minimum of 10 farms, however, several systems were not able to meet this goal and therefore have less than 10 farms.

**Performance metrics**

MCC presents the correlation between the observed outcomes in the data and predicted classifications provided by the model. The value can range from 0 to 1, with values closer to 1 indicating a higher degree of correlation (Chicco and Jurman, 2020; Chicco et al., 2021). Specificity indicates the model's ability to correctly classify individuals as having reported no outbreaks, while sensitivity indicates the algorithm's ability to correctly classify individuals as having reported at least one outbreak. These values, at different classification thresholds, can be used to construct a receiver operating curve from which the AUC can be calculated, to indicate the overall accuracy of the algorithm (Mandrekar, 2010).